\documentclass[10pt]{article}

\usepackage[utf8]{inputenc}
\usepackage{graphicx}
\usepackage{newtxtext,newtxmath}
\usepackage{hyperref}
\hypersetup{
    colorlinks=true,
    linkcolor=black,
    citecolor=black,
    urlcolor=black,
}

\usepackage[a4paper]{geometry}
\newcommand{\ket}[1]{| {#1} \rangle}
\newcommand{\bra}[1]{\langle {#1} |}
\renewcommand{\Re}{\mathrm{Re} \,}
\renewcommand{\Im}{\mathrm{Im} \,}
\newcommand{\tr}{\mathrm{Tr} \,}
\newcommand{\eqnref}[1]{Eq.~(\ref{#1})}
\newcommand{\ii}{\mathrm{i}}
\newcommand{\ee}{\mathrm{e}}

\usepackage{scicite}

\usepackage{authblk}

\title{Hot Schrödinger Cat States} 
\author[1,2,$\dag$]{Ian Yang}
\author[2,3,$\dag$]{Thomas Agrenius}
\author[1,2]{Vasilisa Usova}
\author[2,3,4,5]{Oriol Romero-Isart}
\author[1,2,*]{Gerhard Kirchmair}
\affil[1]{Institute for Experimental Physics, University of Innsbruck, 6020 Innsbruck, Austria}
\affil[2]{Institute for Quantum Optics and Quantum Information, Austrian Academy of Sciences, 6020 Innsbruck, Austria}
\affil[3]{Institute for Theoretical Physics, University of Innsbruck, 6020 Innsbruck, Austria}
\affil[4]{ICFO - Institut de Ciencies Fotoniques, The Barcelona Institute of Science and Technology, Castelldefels (Barcelona) 08860, Spain}
\affil[5]{ICREA, Passeig Lluis Companys 23, 08010, Barcelona, Spain}
\affil[$\dag$]{IY and TA contributed equally to this work.}
\affil[*]{To whom correspondence should be addressed; E-mail: gerhard.kirchmair@uibk.ac.at}
\date{}

\begin{document} 

\maketitle

\begin{abstract}
The observation of quantum phenomena often necessitates sufficiently pure states, a requirement that can be challenging to achieve. In this study, our goal is to prepare a non-classical state originating from a mixed state, utilizing dynamics that preserve the initial low purity of the state. We generate a quantum superposition of displaced thermal states within a microwave cavity using only unitary interactions with a transmon qubit. We measure the Wigner functions of these ``hot'' Schrödinger cat states for an initial purity as low as 0.06. This corresponds to a cavity mode temperature of up to 1.8 Kelvin, sixty times hotter than the cavity's physical environment. Our realization of highly mixed quantum superposition states could be implemented with other continuous-variable systems e.g. nanomechanical oscillators, for which ground-state cooling remains challenging.
\end{abstract}

\section*{Introduction}
The quantum superposition principle allows us to prepare a system in a superposition of two arbitrary states. The paradigmatic example is the superposition of two coherent states, which are pure states with Heisenberg-limited quantum fluctuations~\cite{glauber_Phys.Rev._131:2766_1963}. While the superposition of coherent states is typically called a Schrödinger cat state, in Schrödinger's original thought experiment, the cat — a body-temperature and out-of-equilibrium system — is prepared in a superposition of two mixed states dominated by classical fluctuations \cite{schroedinger_Naturwissenschaften_23:48_1935}.

Experimental demonstrations of Schrödinger cat states typically focus on a continuous-variable degree of freedom, such as the one-dimensional motion of a particle in a harmonic potential~\cite{monroe_Sci_272:5265_1996} or a single electromagnetic field mode in a cavity~\cite{brune_PhysRevLett.77.4887_1996}. The cat states are usually prepared by applying a given ``cat state protocol'' to an initial vacuum Fock state $\hat \rho_0  = \ket 0 \bra 0$, created by cooling the system to its ground state. This results in a pure quantum state which can be written as $\ket \alpha + \mathrm{e}^{\mathrm{i}\phi}\ket{-\alpha}$, where $\ket{\pm\alpha}$ are coherent states of the continuous-variable system with complex amplitude $\alpha$ and relative phase $\phi$. We hereafter call this a ``cold'' Schrödinger cat state. It has an emblematic Wigner function exhibiting an interference pattern, see Figure 1A.

One could ask what type of state would be prepared if the same ``cat state protocol'' is applied to an initial thermal state $\hat \rho_\mathrm{T}$ with a finite average thermal excitation number $n_\mathrm{th}$ \cite{zhu_J.Mod.Opt._43:2_1996, huyet_Phys.Rev.A_63:4_2001, jeong_Phys.Rev.Lett._97:10_2006, zheng_Phys.Rev.A_75:3_2007, jeong_Phys.Rev.A_76:4_2007, nicacio_PhysicsLettersA_374:43_2010}. Would these ``hot'' Schrödinger cat states exhibit quantum features given that (i) the purity of the initial state   $\mathcal{P} = \mathrm{tr} [ \hat \rho_\mathrm{T}^2] = 1/(2 n_\mathrm{th} + 1)$ is substantially less than one for large $n_\mathrm{th}$, and (ii) we consider ``cat state protocols'' that do not remove entropy from the system? In other words, can a highly mixed state exhibit unambiguous quantum features? 

In this article, we experimentally show that indeed, these ``hot'' Schrödinger cat states exhibit quantum features despite being highly mixed. More precisely, we implement two unitary protocols, previously used to prepare cold cats \cite{eickbusch_Nat.Phys.18:12_2022, leghtas_Phys.Rev.A_87:4_2013}, on initial states with a nonzero $n_\mathrm{th}$. We vary $n_\mathrm{th}$ of the initial states up to $7.6(2)$ (corresponding to $\mathcal{P} = 0.062(2)$) and perform direct Wigner function measurements on the final states. The created ``hot'' Schrödinger cat states show Wigner-negative interference patterns for all investigated values of $n_\mathrm{th}$ in the initial state (Figure 1B-G). 

\section*{Results}

Our experimental platform is a circuit quantum electrodynamics (cQED) setup \cite{Reagor_PhysRevB_94:1_2016}. The ``hot'' cat states are prepared in a microwave cavity mode, which is well-described as a quantum harmonic oscillator. The cavity mode is coupled to a two-level system which is used to prepare the cat states. The setup is placed inside a dilution refrigerator and cooled to a temperature of $30$ mK (see Figure S1 for the full experimental schematic). The cavity is a high-coherence $\lambda/4$ post cavity, made up of high-purity niobium, with a resonance frequency $\omega_{c}/2\pi = 4.545$ GHz \cite{heidler_Phys.Rev.A_16:3_2021}, and a relaxation time $T_{1,c}=110(2)$ µs. For the two-level system, we use a transmon qubit with a resonance frequency $\omega_{q}/2\pi =5.735$ GHz, qubit lifetime  $T_{1} = 31.0(4)$ µs, and coherence time $T_2^*= 12.5(4)$ µs. The cavity-qubit interaction is dispersive with the dominant Hamiltonian term $\hat H = -\hbar\chi_\mathrm{qc} \hat c^\dag \hat c \ket e \bra e$. Here, $\hat c^\dag$ and $\hat c$ are the cavity mode creation and annihilation operators, $\ket e$ is the excited state of the qubit, $\chi_\mathrm{qc}/2\pi=1.499(3)$ MHz is the dispersive shift, and $\hbar$ is the reduced Planck's constant. Qubit state measurements are performed through an additional dispersively coupled microstrip resonator with frequency $\omega_{r}/2\pi = 7.534$ GHz. This setup allows for direct measurements of the cavity Wigner function $W(\beta) \equiv 2\, \langle\hat D(\beta)\hat\varPi \hat D^\dagger(\beta)\rangle/\pi$ \cite{vlastakis_Sci_342:6158_2013}. Here $\beta$ is a complex parameter, $\hat D(\beta)$ is the cavity displacement operator, and $\hat\varPi$ is the cavity parity operator \cite{royer_Phys.Rev.A_15:2_1977}. We calibrate the Wigner function measurement by preparing a single photon Fock state in the cavity (Figure S2).

The thermal initial state of the cavity mode, hotter than its environment, is created by equilibrating the cavity mode with a heat bath in the form of filtered and amplified Johnson-Nyquist noise from a 50 ohm resistor. The heat bath is then disconnected (to prevent it from causing additional decoherence) and the cat state preparation commences immediately. The state preparation and measurement protocols take up to $1.9$ µs which is much faster than the cavity relaxation time $T_{1,c}$. Thus, there is neither cooling nor heating of the cavity mode during the protocols. (In Figure S4, we additionally report results from running the experiment with the heat bath left connected during the entire protocol). To verify that the produced initial state is thermal, we characterize the photon statistics of the cavity state with the added noise via qubit spectroscopy \cite{schuster_Nat_445:7127_2007}, which also allows us to relate the noise power to $n_\mathrm{th}$ (Figure S5).

The two protocols used to prepare the hot cat states from the thermal initial state are adaptations of two protocols known in the cQED community as the echoed conditional displacement (ECD) \cite{eickbusch_Nat.Phys.18:12_2022} and qcMAP \cite{leghtas_Phys.Rev.A_87:4_2013} protocols. The quantum circuit diagrams for the protocols we use are shown in Figure~2A. To illustrate the state generation, we decompose the initial thermal state $\hat\rho_T$ in a basis of cavity coherent states $\ket \gamma = \hat D(\gamma)\ket 0$ with $\gamma$ a complex parameter. This lets us discuss the action of the protocols on the state $\ket \gamma \ket g$, where $\ket g$ is the qubit ground state, and then obtain the protocol's action on the actual initial state $\hat\rho_T \otimes \ket g\bra g$ by averaging over $\gamma$ in the explicit decomposition relation $\hat\rho_T = (\pi n_\mathrm{th})^{-1}\int\mathrm{d}^2\gamma \, \exp\{-|\gamma|^2/n_\mathrm{th}\}\ket\gamma\bra\gamma$ \cite{glauber_Phys.Rev._131:2766_1963}. A pictorial version of this argument represents the thermal state as a ``compass'' of four slightly displaced coherent states, $\gamma = \{\epsilon, \ii\epsilon, -\epsilon, -\ii\epsilon\}$ for a small $\epsilon > 0$, and tracks their motion in phase-space during the protocol (Figure 2B,C). The first operation sequence prepares the qubit in the superposition state $\ket g + \ee^{\ii\phi}\ket e$, where $\phi$ is a controllable phase shift, and the cavity in the displaced state $\ket{\gamma + \alpha}$.
Next, the cavity-qubit state becomes entangled through time evolution under the dispersive interaction $\hat H$, as illustrated in Figure 2B and C for ECD and qcMAP respectively. The qcMAP protocol has an uninterrupted time evolution, while the ECD protocol has additional displacements and a qubit echo pulse inserted at half the evolution time. At the end of the time evolution, we have created the state $\ee^{\ii\phi}\ket {\ii\gamma - \alpha}\ket g + \ket {\ii\gamma + \alpha}\ket e$ with the ECD protocol (Figure 2B) and the state $\ket {\gamma + \alpha}\ket g + \ee^{\ii\phi}\ket {- \gamma - \alpha}\ket e$ with the qcMAP protocol (Figure 2C). The final three operations in Figure 2A disentangle the qubit from the cavity. We center the $\ket e$ branch with a displacement $-\alpha$ for ECD and $\alpha$ for qcMAP. Next, we apply a qubit $\pi$-pulse selective to the $\ket e$ branch only. Finally, we invert the previous displacement. The selective $\pi$-pulse has a Gaussian envelope with standard deviation in time $\sigma_t$. This induces a Fock number-dependent qubit flip probability $P_{g\leftrightarrow e}(n) = \exp\{-(\chi_\mathrm{qc}\sigma_t n)^2\}$ \cite{thomas_Phys.Rev.A_27:5_1983, mihov_PulseShape_2023}. In contrast to the known qcMAP and ECD protocol, where only the qubit state that is entangled with the cavity zero Fock state is affected, here we choose $\sigma_t$ so that this probability is large for the non-displaced thermal state but small for the displaced thermal state (Figure 2D). This requires that $\alpha$ is chosen large enough so that the Fock number distributions of the initial thermal state and the thermal state displaced by $2\alpha$ do not overlap (roughly, $|\alpha|^2 > n_\mathrm{th} + 1$). In a phase-space picture, we flip the qubit only for phase-space points within a finite radius of the origin (Figure 2E).

We run our experiment with an initial thermal excitation number of $n_\mathrm{th} = 0.75(1)$, $1.44(2)$, $1.84(3)$, $3.48(7)$, $7.6(2)$, corresponding respectively to purities of $\mathcal{P}=0.400(3)$, $0.258(3)$, $0.214(3)$, $0.126(2)$, $0.062(2)$. We use $\alpha=3$, set $\phi=0$, and use a disentanglement pulse width of $\sigma_t = 20$ ns. Figure 1B,C show Wigner function measurements on a grid of phase-space points $\beta$ for $n_\mathrm{th} = 3.48$. Figure 1D-G show Wigner function measurements along the $\Re\{\beta\}$ and $\Im\{\beta\}$ axes as $n_\mathrm{th}$ is varied. Figure 1H-J show Fock density matrices reconstructed from the Wigner measurements in Figure 1A-C; see Materials and Methods for the reconstruction method. In Supplemental Figure S3, we additionally show Wigner function measurements for $n_\mathrm{th} = 1.84(3)$. While the ECD and qcMAP protocols are known to prepare equivalent cold cats, we observe that they lead to distinct outcomes when applied to thermal initial states; compare panels B and C in Figure 1. In both cases, the Wigner functions have two separated Gaussians centered at $\alpha$ and $-\alpha$, associated with the classical probability distribution of the displaced thermal initial states. Centered between these is an interference pattern of oscillations with negative values, which produces an interference pattern in the marginal distribution along $\Im\{\beta\}$. In the ECD state, the envelope of the interference pattern grows in radius and decreases in amplitude with $n_\mathrm{th}$ similar to the displaced thermal states themselves (Figure 1D,E). In the qcMAP state, the envelope shrinks with increasing $n_\mathrm{th}$, but its amplitude decreases more slowly than the displaced thermal states (Figure 1F,G). The data for all prepared states show clear negative values in the interference pattern regardless of their $n_\mathrm{th}$.

These results can be understood as follows. Under ideal conditions, the ECD and qcMAP protocols are equivalent to the application of two different quantum operators, namely $\hat S_{1} \equiv [\hat D(\alpha) + \mathrm{e}^{\mathrm{i}\phi}\hat D(-\alpha)]/\sqrt{2}$ (ECD) and $\hat S_{2} \equiv [1 + \mathrm{e}^{\mathrm{i}\phi} \hat\varPi ]\hat D(\alpha)/\sqrt{2}$ (qcMAP), to the cavity initial state $\hat\rho_T$ . (The ECD protocol is more generally equivalent to the operator $\hat S_{1}' \equiv \left.\hat S_1\right|_{\phi \to \phi + 2|\alpha|^2} \mathrm{i}^{\hat c^\dagger \hat c}$, but these differences vanish when the initial state is thermal and the geometric phase $2|\alpha|^2$ is cancelled by $\phi$; see Materials and Methods). The Wigner functions prepared from $\hat\rho_T$ by the operators $\hat S_{1,2}$ are $W_{1,2}(\beta) \equiv \frac 1 2 \left[W_T(\beta - \alpha) + W_T(\beta + \alpha) + C_{1,2}(\beta)\right]$ where $W_T(\beta \pm \alpha)$ are the Wigner functions of the left- and right-displaced thermal states, and the third term, which represents the quantum superposition property of the state, is different for the two preparations: $C_{1,2}(\beta) = 2\cos\left(4\Im\{\alpha^*\beta\} +\phi\right)f_{1,2}(\beta)$.
For ECD, $f_1(\beta) \equiv W_T(\beta) = 2\mathcal{P}\mathrm{e}^{-2\mathcal{P}|\beta|^2}/\pi$ is the thermal initial state. For qcMAP, $f_{2}(\beta) \equiv 2\mathrm{e}^{-2|\beta|^2/\mathcal{P}}/\pi$ is related to the characteristic function~\cite{barnett_MethodsTheoretical_2002} of the initial state. We plot examples of $W_{1,2}(\beta)$ in Figure 3A.
As illustrated there, when $n_\mathrm{th} \to 0$ ($\mathcal{P} \to 1$), $f_{1,2}(\beta)$ become identical and $W_{1,2}(\beta)$ both become equal to the cold cat Wigner function. When $n_\mathrm{th} > 0$, the phase-space radius of $f_1(\beta)$ grows at the same rate as the phase-space radii of $W_T(\beta \pm \alpha)$, while the phase-space radius of $f_2(\beta)$ shrinks. For both states, the $\Im\{\beta\}$ marginal probability distribution contains interference fringes with period $\pi/2\alpha$ and full contrast independently of $n_\mathrm{th}$. We remark that for $\phi = n\pi$ with $n$ integer, $W_{2}(0)$ is always saturated to the Wigner function upper/lower bounds $\pm 2/\pi$, corresponding to parity values $\langle \hat\varPi \rangle = \pm 1$, independently of $n_\mathrm{th}$. Realizing this saturation of parity could be useful for hardware-efficient encoding in bosonic qubit states in the presence of finite mode temperature. For example, in \cite{grimm_Nat._584:7820_2020, lescanne_Nat.Phys._16:5_2020}, the stabilization process via two-photon dissipation or parametric pumping is parity preserving. An initial thermal cavity state will reduce the prepared state fidelity. This can be overcome by first using the qcMAP operator before stabilization. The sharp saturation of parity in a small phase-space volume in the qcMAP state could also be explored for applications in quantum metrology. For comparison with Figure 1H-J, we display the Fock density matrices of the ideal hot cat states in Figure 3B.

To highlight the quantum superposition nature of the hot cat states, we consider the coherence function $g(x_1,x_2) \equiv |\bra{x_1}\hat\rho\ket{x_2}|/\sqrt{\bra{x_1}\hat\rho\ket{x_1}\bra{x_2}\hat\rho\ket{x_2}}$ of the prepared states. Here, $x_{1,2}$ are eigenvalues and $\ket{x_{1,2}}$ eigenkets of the dimensionless quadrature operator $\hat x \equiv (\hat c + \hat c^\dagger)/\sqrt 2$, which is defined in analogy to the position operator of a mechanical harmonic oscillator. The coherence function compares the off-diagonal elements of the density matrix to their maximum possible value, so is upper-bounded as $g(x_1,x_2) \leq 1$. 
The value of $g(\sqrt 2 \alpha, -\sqrt 2 \alpha)$ quantifies, on a continuous scale from 0 to 1, whether the oscillator is observed to be in a quantum superposition of being displaced by $\alpha$ and $-\alpha$ (the $\sqrt 2$ factor is due to the definition of the position quadrature operator $\hat x$). This value is also directly related to the contrast of the interference fringes in the marginal distributions \cite{glauber_Phys.Rev._130:2529_1963}. We reconstruct the coherence functions of our experimentally prepared states along the line from $(x_1,x_2) = \sqrt{2}\alpha(1, 1)$ to $\sqrt 2 \alpha(1, -1)$ via the reconstructed density matrices of Figure 1H-J. We present the results in Figure 4 along with theoretical curves. Consider first the ideal case, represented by solid lines in Figure 4. For both states prepared by $\hat S_{1,2}$, $g(x_1,x_2) = \ee^{-(|x_1|-|x_2|)^2/2\xi_\mathrm{th}^2}$ along the considered line. (Away from this line, the coherence functions of $W_{1,2}(\beta)$ are different; see Supplemental Figure S8). The quantity $\xi_\mathrm{th} = \sqrt{2\mathcal{P}/(1-\mathcal{P}^2)} \approx 1/\sqrt{n_\mathrm{th}}$ is the thermal coherence length. When $x_1 \approx x_2$, the cat state coherence functions behave as those of thermal states, $\ee^{-|x_1-x_2|^2/2\xi_\mathrm{th}^2}$, with the corresponding values of $n_\mathrm{th}$. As $n_\mathrm{th}$ increases, $\xi_\mathrm{th}$ decreases and the thermal state coherence function decays faster with $|x_1 - x_2|$ (Figure 4A). The peak at $x_1 = - x_2 = \sqrt 2\alpha$ in the cat state coherence functions (solid lines in Figure 4B-D), which is not present in the thermal state coherence function, indicates the quantum superposition nature of these states. In the ideal case, this peak saturates to the upper bound $g(\sqrt 2 \alpha, -\sqrt 2 \alpha) = 1$, with the value of $n_\mathrm{th}$ only affecting how narrow this extra peak is. The data (dash-dotted lines in Figure 4B-D) shows similar behavior to the ideal case, but deviates both due to the limitations of the density matrix reconstruction method as well as due to the differences between the experiment and the ideal case. We discuss the latter further below. Here, we note that the cat state data always displays a peak around $(x_1,x_2) = \sqrt 2\alpha(1,-1)$, the height of which is not substantially affected by the value of $n_\mathrm{th}$.

\section*{Discussion}

We emphasize that the additional peak in the cat state coherence function comes from the  unitary operations applied to the initial state (i.e.~the cat creation operators $\hat S_{1,2}$). These operations thus generate coherence not present in the initial state. Both protocols generate the same amount of coherence (as shown by the coherence function), resulting in equally coherent superpositions of mixed states, despite the parity of the hot ECD cat state being lower than the hot qcMAP state (as shown by the Wigner functions at the origin $W_{1,2}(0)$).
The additional peak in the coherence function is also generated in experiments which observe interference patterns from superpositions of thermal clouds of atoms prepared using Bragg diffraction \cite{miller_Phys.Rev.A_71:4_2005} or Stern--Gerlach interferometry \cite{margalit_NewJ.Phys._21:7_2019}. In contrast, the additional peak is not present in the state obtained by sending a thermal state through a double slit grating (see e.g.~\cite{bloch_Nature_403:6766_2000}), nor in a completely dephased cold cat state, namely $\hat \rho_\mathrm{dephased} = \frac{1}{2}(\ket\alpha \bra \alpha + \ket{-\alpha}\bra{-\alpha})$ (see e.g.~\cite{deleglise_Nature_455:7212_2008}). Note that $\hat \rho_\mathrm{dephased}$ has purity 1/2 but a completely positive Wigner function ($W_{1,2}(\beta)$ with $C_{1,2}(\beta) = 0$), whereas we experimentally prepare states with negative Wigner functions down to purities of 0.06.

The measured Wigner functions (Figure 1) deviate from the ideal Wigner functions $W_{1,2}(\beta)$ (Figure 3A) mainly due to qubit operation imperfections, perturbative Hamiltonian nonlinearities and cavity-qubit system decoherences. Through the reconstructed density matrices, we estimate the fidelity of the experimentally prepared states to the ideal cat states $W_{1,2}(\beta)$ to be 0.8 for the cold cat and 0.7 for the hot cats. The loss in fidelity can be partly accounted for by qubit coherence time (3\%), cavity lifetime (5\%) and qubit population (2\%). To determine the role of other factors, we use our characterization of the experiment (Figure S6, Table S1) to construct an \emph{ab initio} model of the state preparation protocols that includes the nonlinearities, the shape and timings of the qubit pulses, and decoherence during state preparation. The Wigner function measurement is modelled as a projective measurement that includes the effects of residual cavity-qubit entanglement due to pulse imperfections (see Materials and Methods). We find that the model reproduces most of the imperfections seen in the data, see Figures S9-S10. The main remaining discrepancy is that the model states are more coherent than the reconstructed states, as can seen by comparing dashed and dash-dotted lines in Figure~4. In this comparison, the model states are expected to show only half the decoherence when compared to the reconstructed states, since the model excludes decoherence during the measurement sequence. The remaining extra decoherence is due to uncharacterized additional loss channels, most likely cavity dephasing, whose presence is indicated by cat parity lifetime measurements (Figure~S7). We leave the exact characterization of this noise for future work. Adding a small amount of dephasing to the \emph{ab initio} model improves the fidelity to the reconstructed states as well as the qualitative agreement (for illustration, see the dotted curve in Figure 4 which has an added cavity dephasing rate of $1/(80\ \text{µs})$).

By running the \emph{ab initio} model and toggling each feature of the model, such as operation imperfections or Hamiltonian nonlinearities, between being included and excluded, we attribute each discrepancy between the measured and ideal Wigner functions to a cause (Figures S11-S13). Here, we summarize the conclusions and further details are given in Supplemental Materials section S4. The leading perturbative Hamiltonian terms $\hat H' = - \frac 1 2(K_\mathrm{c} + \chi'_\mathrm{qc}\ket e \bra e)\hat c^\dag \hat c^\dag \hat c \hat c$ ($K_\mathrm{c}/2\pi = 4.9(1)$ kHz, $\chi'_\mathrm{qc}/2\pi = 12.8(9)$ kHz) cause a smearing and bending distortion of the Wigner functions which is similar to that observed in previous experiments \cite{kirchmair_Nat_495:7440_2013}. The 20 ns width of the disentanglement pulse causes left-right asymmetries of the displaced thermal states as well as bending distortions of the fringes. In particular, the nonlinearities and finite pulse widths together cause the qcMAP linecuts (Figure 2D) to deviate from those predicted by $W_2(\beta)$, with an $n_\mathrm{th}$-dependent reduction of the maximum of $C_2(\beta)$. The extra fringes in the ECD state Wigner function as compared to $W_1(\beta)$ are caused by instrumentation limitations to the minimum qubit pulse width that we can achieve (Gaussian standard deviation of 6 ns).

We have demonstrated and characterized the preparation of quantum superposition states directly from thermally excited initial states using only unitary dynamics. Preparing ``hotter'' (larger initial thermal occupation number) cat states in our setup requires using larger cavity displacements so that the ancillary qubit can be disentangled from the cavity in the final step of the protocol. Limitations eventually appear due to the finite coherence time, finite pulse width, and perturbative nonlinearities in our experiment.  In this context, we note that state-of-the-art cQED setups capable of cold cat states with $\alpha = 32$ were recently reported \cite{Milul_PRXQuantum.4.030336_2023}. In other systems, limitations arise in the measurement of the prepared state rather than in its preparation. For example, the narrow thermal coherence length of the hot cats leads to increased time-of-flight requirements to see interference fringes in a Bragg~\cite{miller_Phys.Rev.A_71:4_2005} or half-Stern--Gerlach interferometer~\cite{margalit_NewJ.Phys._21:7_2019} and leads to increased recombination precision requirements in a full-loop Stern--Gerlach interferometer~\cite{margalit_Sci.Adv._7:eabg2879_2021}, compared to cold cats. Nevertheless, under ideal conditions including ideal measurement precision, standard quantum-mechanical theory predicts no upper limit on and no loss of contrast due to the thermal occupation number of a hot cat state \cite{zhu_J.Mod.Opt._43:2_1996, huyet_Phys.Rev.A_63:4_2001, jeong_Phys.Rev.Lett._97:10_2006, zheng_Phys.Rev.A_75:3_2007, jeong_Phys.Rev.A_76:4_2007, nicacio_PhysicsLettersA_374:43_2010}.
Hot Schrödinger cat states are in principle realizable in any continuous-variable quantum system. This is particularly relevant for systems where long coherence times have been achieved but ground-state cooling is not (yet) available. Specific examples include nanomechanical systems such as carbon nanotubes \cite{samanta_Nat.Phys._19:9_2023}, and levitated magnetic \cite{gutierrezlatorre_Phys.Rev.Appl._19:054047_2023, hofer_Phys.Rev.Lett._131:043603_2023} and electrostatically trapped dielectric particles \cite{millen_Phys.Rev.Lett._114:123602_2015, delord_Nature_580:56_2020, conangla_NanoLett._20:6018_2020, dania_Phys.Rev.Lett._132:133602_2024}. Our work highlights both opportunities and challenges associated with designing protocols for the observation of quantum phenomena that do not require ground-state cooling \cite{scala_Phys.Rev.Lett._111:18_2013, Marshall_PhysRevA.99.032345_2019}.

\section*{Materials and Methods}

\subsection*{Fabrication}

\noindent The transmon qubit and readout resonator were patterned by electron-beam lithography (Raith eLINE Pllus $30$ kV) on a bi-layer resist ($1$ µm MMA (8.5) EL13 and $0.3$ µm of 950 PMMA A4). The substrate started from a 2-inch sapphire wafer that was first piranha-cleaned before processing. To prevent charging of the substrate, a thin gold layer was sputtered on top of the PMMA. After lithography, this gold layer was etched in a solution of Lugol (5 \% potassium iodide) and DI water in a ratio of 1:15, before being washed in DI water and developed in a 3:1 solution of isopropyl alcohol and water. In the next step, two layers of aluminum ($25$ nm and $50$ nm) were evaporated onto the sample using a Plassys MEB550S electron-beam evaporator. A controlled oxidation step ($5$ mbar for $5.5$ min) was carried out in between the deposition of the two aluminum layers. Subsequently, the qubit chip was laser-diced, and the resist layer was lifted off. The sample chips were thermalized by a copper clamp. More details about the experimental setup can be found in the supplementary materials section S1.

\subsection*{Density matrix and coherence function reconstruction and fidelity estimates}

\noindent 
To reconstruct the density matrices $\hat\rho$ of the prepared states, we compute the matrix elements in the Fock basis $\rho_{mn}$ as
$    \rho_{mn} = \bra{m} \hat\rho \ket {n} = 2\int\mathrm{d}^2\beta\, W(\beta) \bra{m}\hat\varPi(\beta)\ket{n}
$
where $\hat\varPi(\beta) = \hat D(\beta)\hat \varPi \hat D^\dagger(\beta)$ and $\ket n$ is the $n$:th cavity Fock state.
We approximate the integral as a Riemann sum on the phase-space grid of measured data for $W(\beta)$. The matrix $\rho_{mn}$ that we obtain from the data is not a density matrix, but rather a noisy estimate of the operator $\hat\rho = p_g\hat\rho_{g} - p_e\hat\rho_{e}$, where $p_e$ is the residual excited state qubit population after the state preparation protocol, $p_g = 1-p_e$, and $\hat \rho_{g}$ ($\hat \rho_{e}$) is the cavity state conditional on the qubit being in the ground (excited) state. The state we are seeking to reconstruct is $\hat \rho_g$. To extract the corresponding Fock matrix $\rho_{g}$ from the matrix $\rho = (\rho_{mn})$, we diagonalize $\rho$ and then construct $\rho_{g}$ as the positive semidefinite part of $\rho$. From the reconstructed density matrix $\rho_g$, we compute the fidelity $\mathcal{F}$ of the considered state to a reference state $\hat\rho'$ as $\mathcal{F} = \left(\tr\left\{\sqrt{\sqrt{\hat\rho'}\hat\rho_g\sqrt{\hat\rho'}}\right\}\right)^2$. We compute the coherence function by expressing $\hat\rho_g$ in the eigenbasis of $\hat x$. Formulas and further details of the steps described here are given in the Supplementary Materials section S1.6.

\subsection*{Derivation of equivalent operators for the ideal pulse sequences and the ideal Wigner functions}

\noindent Here, we summarize a longer derivation that we give in Supplementary Materials sections S2.1-2. The ECD and qcMAP protocols are described in Hilbert space by two unitary operators $\hat U_{1,2}$ which are given in the circuit diagrams of Figure~2. When applied to an initial state $\hat\rho_0\otimes\ket g \bra g$, where $\hat\rho_0$ is a general initial cavity state, the protocols result in total cavity-qubit states $\hat\rho_f = \hat S_{j,gg}\hat\rho_0\hat S^\dagger_{j,gg} \otimes \ket g \bra g + \hat S_{j,eg} \hat\rho_0 \hat S_{j,eg}^\dagger \otimes \ket e \bra e + \hat\psi$, where $\hat S_{j,gg} = \bra g \hat U_j \ket g$ and $\hat S_{j,eg} = \bra e \hat U_j \ket g$ for $j\in\{1,2\}$, and $\hat\psi$ represents coherence terms between $\ket g$ and $\ket e$. We identify $p_g = \tr\{\hat S_{j,gg}\hat\rho_0\hat S^\dagger_{j,gg}\}$, $\hat\rho_g = \hat S_{j,gg}\hat\rho_0\hat S^\dagger_{j,gg}/p_g$, and similar expressions for $p_e$ and $\hat\rho_e$, so that $\hat\rho_f = p_g\hat\rho_g \otimes \ket g\bra g + p_e\hat\rho_e \otimes \ket e\bra e + \hat\psi$. Unitarity of $\hat U_{1,2}$ implies $p_g + p_e = 1$. If $p_g = 1$, then acting with $\hat U_j$ on the cavity-qubit system effectively applies the operator $\hat S_{j,gg}$ to the cavity initial state while leaving the qubit in $\ket g$. From $\hat U_{1,2}$ one derives
$\hat S_{1,gg} = \frac{1}{\sqrt 2}\hat D(\alpha)\left[\ii\ee^{\ii(\phi + |\alpha|^2)} \bra g \hat X(\pi, \sigma_t) \ket g \hat D(-2\alpha) - \ee^{-\ii|\alpha|^2}\bra g \hat X(\pi, \sigma_t) \ket e\right]\ii^{\hat c^\dagger\hat c}$,
$\hat S_{2,gg} = \frac{1}{\sqrt 2}\hat D(-\alpha)\left[\bra g \hat X(\pi, \sigma_t) \ket g\hat D(2\alpha) + \ii\ee^{\ii\phi} \hat\varPi \bra g\hat X(\pi, \sigma_t) \ket e\right]$,
where $\bra g \hat X(\pi, \sigma_t) \ket e$ and $\bra g \hat X(\pi, \sigma_t) \ket g$ are operators on the cavity Hilbert space. 
If $\bra g \hat X(\pi, \sigma_t) \ket e \hat\rho_0 = \hat\rho_0$ while $\bra g \hat X(\pi, \sigma_t) \ket g\hat D(\pm 2\alpha)\hat\rho_0 = \hat D(\pm 2\alpha)\hat\rho_0$, then it follows that $p_g = 1$. A necessary condition is that $\hat\rho_0$ has no overlap with itself when displaced by $2\alpha$. We then recover $\hat S_{1}'$ and $\hat S_{2}$ from $\hat S_{1,gg}$ and $\hat S_{2,gg}$ after dropping global phases and redefining $\phi$ to cancel relative phases. The stated conditions are the formal versions of the condition that the disentanglement pulse should flip the qubit state only for the cavity state entangled to $\ket e$. The ideal Wigner functions $W_{1,2}(\beta)$ follow directly from the Wigner representations of $\hat\rho_g$ when $p_g = 1$.

\subsection*{Numerical model}

\noindent We simulate the ECD and qcMAP protocols using QuTiP version 4.7 \cite{qutip}. We model the time evolution operations $\hat T$ and the qubit and cavity-conditional qubit operations $\hat X, \hat Y$ by numerically integrating the master equation describing the total cavity-qubit dynamics with parameters (system Hamiltonian, decay rates, qubit driving Hamiltonian) as characterized in the experiment. Displacement operators and thermal initial states are implemented using QuTiP’s built-in functions. We model the Wigner function measurement as the expectation value of the observable
$\hat M(\beta) \equiv \frac{2}{\pi}\hat\varPi(\beta)\left(\ket g \bra g - \ket e \bra e\right)$ in the final states of the protocols. This model accounts for the effect of residual qubit $\ket e$ population after state preparation but neglects the influence of decoherence and non-linearities during the measurement sequence \cite{vlastakis_Sci_342:6158_2013}.
 More details are given in Supplementary Materials section S3. Details on the subsequent investigation to study the effects of operation imperfections and Hamiltonian nonlinearities individually are given in section S4. 

\nocite{horowitz_textbook_CUP_1989, Diedrich_ion_sideband_cooling_1989, Hamann_atom_sideband_cooling_1998, smolin_Phys.Rev.Lett._108:070502_2012, qutip, kim_Phys.Rev.A_46:7_1992}
\bibliography{references}
\bibliographystyle{sciencemag_modified}
\clearpage

\section*{Acknowledgments}

TA and ORI wish to thank Lukas Neumeier for useful discussions. 
IY, VU and GK wish to thank Johannes Fink for the etching of our Niobium cavities.

\textbf{Funding:}
IY, VU and GK were funded in part by the Austrian Science Fund (FWF) DOI 10.55776/F71 and 10.55776/I4395 QuantERA grant QuCOS. For the purpose of open access, the author has applied a CC BY public copyright licence to any Author Accepted Manuscript version arising from this submission.
TA and ORI were supported by the European Research Council (ERC) under Grant Agreement No.~[951234] (Q-Xtreme ERC-2020-SyG).

\textbf{Author contributions:} IY carried out the experiment with support from VU. TA carried out the theoretical analysis with support from the other authors. The numerical simulations were done by IY, TA, and VU. ORI and GK conceived of and supervised the project. All authors were involved in the writing and the editing of the paper.

\textbf{Competing interests:} The authors declare no competing interests.

\textbf{Data availability:} The data that supports the plots within this paper and the supplementary material is available via a permanent, public repository, Zenodo https://doi.org/10.5281/zenodo.13960693.

\clearpage

\section*{Figures and Tables}

\makebox[\textwidth]{\includegraphics{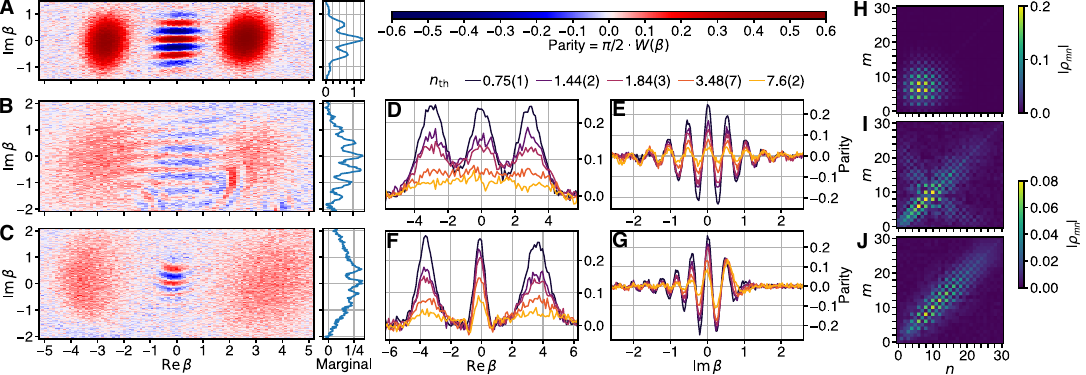}}
\noindent {\bf Fig. 1. Wigner function measurement results.} {\bf (A)} Cold Schrödinger cat prepared using the qcMAP protocol with the heat bath disconnected ($n_\mathrm{th} = 0.03$, $\mathcal{P} = 0.94$). {\bf (B,C)} `Hot' Schrödinger cat states prepared from an initial thermal state with $n_\mathrm{th} = 3.48(7)$ using the ECD (B) and qcMAP (C) protocols. Also displayed in A-C are the marginal distributions obtained by summation along the $\Re\{\beta\}$ axis. To increase the visibility of small parity values, the color brightness changes nonlinearly across the colorbar. {\bf (D-G)} Linecuts of the Wigner function along the coordinate axes in the ECD (D,E) and qcMAP (F,G) protocols with $n_\mathrm{th}$ of the initial state as indicated in the legend. (\textbf{H,I,J}) Fock density matrices reconstructed from the Wigner measurement data in panel A,B,C respectively. Panels I and J share the lower colorbar. The matrix elements are given by $\rho_{\mathrm{mn}} = \bra{m} \hat\rho \ket {n}$.
\clearpage

\makebox[\textwidth]{\includegraphics{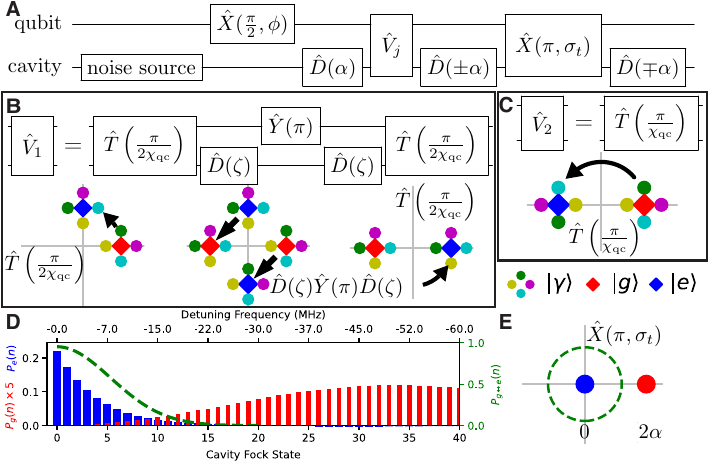}}
\noindent {\bf Fig. 2. Hot cat state generation protocol.}  {\bf (A)} Quantum circuit diagram of hot cat state generation sequence. For ECD (qcMAP), $j=1$ ($j=2$), and the displacements use the lower (upper) sign. $\hat X(\pi/2, \phi)$ is a qubit $\pi/2$ pulse with phase $\phi$, and $\hat X(\pi, \sigma_t)$ is the disentanglement pulse. {\textbf{(B)}} Definition of $\hat V_1$ and visualization of its action in the joint cavity-qubit phase space. $\hat T(t)$ denotes free evolution for time $t$, $\hat Y(\pi)$ is a qubit $\pi$ pulse, and $\zeta = -(1+i)\alpha/2$. Cavity states are entangled with the qubit states whose marker they touch. The arrows illustrate how the total state evolves under the indicated operator. {\textbf{(C)}} Definition and visualization of $\hat V_2$. {\textbf{(D)}} Qubit-conditional cavity Fock state distributions $P_{q}(n) = \bra n \bra q \hat\rho \ket q \ket n$, ($q\in\{g,e\}$) of the total state before the $\hat{X}(\pi,\sigma_t)$ operation. Here $\alpha = 3$ and $n_{\mathrm{th}} = 3.5$. The dashed line shows $P_{g\leftrightarrow e}(n)$ (defined in the main text) with $\sigma_\mathrm{t} = 20$ ns. {\textbf{(E)}} The choice of $\sigma_t$ corresponds to the choice of a radius in the cavity-qubit phase space within which the qubit state is flipped with a certain probability. At this stage of the protocol, the $\ket g$ branch is displaced by $2\alpha$.
\clearpage

\makebox[\textwidth]{\includegraphics{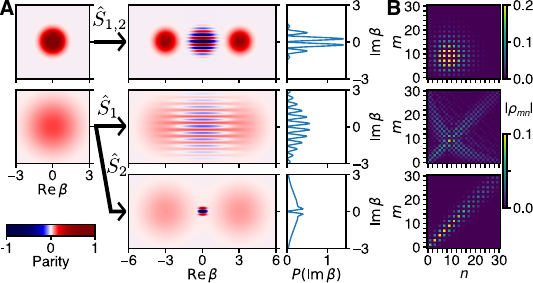}}
\noindent {\bf Fig. 3. Hot Schrödinger Cat States in Theory.} {\bf (A)} Left: Plots of the thermal state Wigner function $W_T(\beta)$ with $n_\mathrm{th} = 0$ (top), $n_\mathrm{th} = 3.5$ (middle). To increase the visibility of the hotter state, the color brightness changes nonlinearly across the colorbar (bottom). Center: Cat state Wigner functions $W_{1,2}(\beta)$ which result from applying the operators $\hat S_{1,2}$ with $\alpha=3.5$ and $\phi = \pi$ to the initial states on the left according to the arrows. Right: Marginal probability distributions obtained from the cat state Wigner functions by integrating along the $\Re\{\beta\}$ axis. {\bf (B)} Absolute values of the Fock density matrix elements corresponding to the hot cat states displayed on the same row. The middle and lower panels share the lower colorbar.  The matrix elements are given by $\rho_{\mathrm{mn}}= \bra{m} \hat\rho \ket {n}$.
\clearpage

\makebox[\textwidth]{\includegraphics{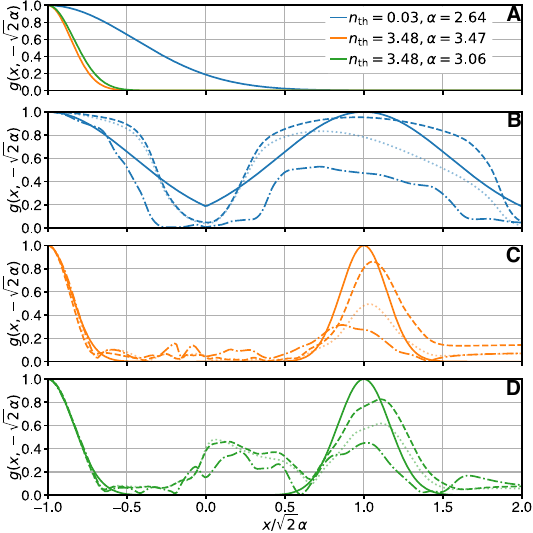}}
\noindent {\bf Fig. 4. Ideal, reconstructed, and numerically simulated coherence functions.} {\bf (A)} Sections of the theoretical thermal state coherence function (given in the text) with values of $\alpha$ and $n_\mathrm{th}$ as indicated in the panel legend. {\bf (B,C,D)} Sections of the coherence function for the (B) cold cat (C) hot qcMAP cat (D) hot ECD cat states along the line from $(x_1,x_2)=\sqrt 2 \alpha(-1,-1)$ to $(x_1,x_2)=\sqrt 2 \alpha(2,-1)$. The solid lines indicate the ideal states with the values of $n_\mathrm{th}$ and $\alpha$ given in the legend in panel A. The dash-dotted line is computed from the density matrix reconstruction of the measured experimental data (Figure 1H-J). Dashed lines are computed from density matrix reconstructions performed on the output of the \emph{ab initio} numerical model of the state preparation discussed in the text. The dotted semitransparent lines are included to demonstrate the output of the numerical model if additional cavity dephasing is added, see Discussion.
\clearpage

\section*{Supplemental Materials}
\newcommand{\braket}[1]{{\langle #1 \rangle}}
\newcommand{\dd}{\mathrm{d}}
\newcommand{\vb}[1]{{\boldsymbol{#1}}}

\renewcommand{\thetable}{S\arabic{table}}
\renewcommand{\thefigure}{S\arabic{figure}}
\renewcommand{\theequation}{S\arabic{equation}}
\renewcommand\thesection{S\arabic{section}}

\section{Experiment}
\subsection{\label{s:setup}Experimental Setup}
The schematic of the experimental setup is shown in Figure~S1. The high coherence cavity has a post length of $14.8$ mm, inner radius of $2$ mm and outer radius of $6.4$ mm. This geometry gives a bare cavity frequency of approximately $4.5$ GHz. The tunnel for the qubit chip has a diameter of $4$ mm, which is a compromise between cavity mode leakage into the tunnel and qubit capacitance to ground. The cavity was made from high-purity niobium at the Institute for Quantum Optics and Quantum Information Innsbruck mechanical workshop. The manufacturing process used electro-discharge machining with a tungsten alloy electrode. The cavity was then etched with our collaborators at the Institute of Science and Technology, Vienna with the group of Prof. Johannes Fink. This process used a buffer chemical polishing etching solution of 1:1:1 hydrofluoric, nitric and phosphoric acid for one hour at 5 $^{\circ}$C. Phosphoric acid was then slowly added to reach a ratio of 1:1:2 for another hour of polishing. Afterwards, the niobium cavity was rinsed heavily with deionized (DI) water. In total, this process removes approximately $150$ µm of material. 

The transmon qubit and readout resonator were patterned by electron-beam lithography (Raith eLINE Pllus $30$ kV) on a bi-layer resist ($1$ µm MMA (8.5) EL13 and $0.3$ µm of 950 PMMA A4). The substrate started from a 2-inch sapphire wafer that was first piranha-cleaned before processing. To prevent charging of the substrate, a thin gold layer was sputtered on top of the PMMA. After lithography, this gold layer was etched in a solution of Lugol (5 \% potassium iodide) and DI water in a ratio of 1:15, before being washed in DI water and developed in a 3:1 solution of isopropyl alcohol and water. In the next step, two layers of aluminum ($25$ nm and $50$ nm) were evaporated onto the sample using a Plassys MEB550S electron-beam evaporator. A controlled oxidation step ($5$ mbar for $5.5$ min) was carried out in between the deposition of the two aluminum layers. Subsequently, the qubit chip was laser-diced, and the resist layer was lifted off. The sample chips were thermalized by a copper clamp. An additional aluminum sheet was used to cover the copper clamp to reduce losses due to the presence of the copper material.

The measurements were conducted in a Triton DU7-200 Cryofree dilution refrigerator system. The input coaxial cables were attenuated by $20$ dB at the $4$ K plate and $10$ dB at the still plate. Finally, at the mixing chamber plate, the input signal was filtered by a K\&L DC-12 GHz low pass filter and then attenuated by a $20$ dB directional coupler followed by a thermalized cryogenic $20$ dB attenuator and filtered by microtronics $4-8$ GHz bandpass filter. The experiment was done in reflection with a Quinstar double junction $4-8$ GHz circulator. Before and after the sample, the input and output signals passed through a home-built eccosorb filter. The input signal for the high coherence cavity was attenuated and filtered similarly, except at the base plate where a $10$ dB thermalized cryogenic attenuator was used instead.

The output signal was filtered via a  microtronics $4-8$ GHz bandpass filter, before passing through a quantum-limited parametric amplifier. Finally, the output signal was filtered by a K\&L filter which was connected to two Quinstar isolators giving $40$ dB isolation. The signal was amplified at the $4$ K plate by high electron mobility transistor (HEMT) amplifiers and again with room temperature amplifiers outside of the refrigerator. 

Control of the thermal noise was done by amplifying and filtering the noise from a 50 ohm resistor. The added noise has a frequency spectrum shown in Figure S4A. The added noise power level was controlled by a digital attenuator. The setup allowed for a maximum of $60$ dB of added thermal noise. A fast, home-built microwave switch was used to disconnect the cavity mode from this added noise. The switch has an open attenuation of $40$ dB and has a rise and fall time of around $10$ ns. To initialize the cavity state, the cavity mode was allowed to come into thermal equilibrium with the controlled noise environment for 1 ms. Afterwards, the microwave switch was opened and the state preparation and measurement started.

Leaving the microwave switch opened resulted in a thermal state in equilibrium with the residual thermal excitations of the setup ($n_\mathrm{th} = 0.0338(7)$), which was the coldest initial state we could achieve with this setup.

The samples were placed in a $\mu$-metal shield which sat in a superconducting shield to protect the experiment against stray magnetic fields. The shield was filled with eccosorb foam for the absorption of any stray infrared photons. 

The pulses for the high coherence cavity and readout resonator were generated by an arbitrary waveform generator (AWG), specifically the Operator X from Quantum Machines. These pulses were up-mixed with a local oscillator (LO) using a Marki microwave IQ mixer. The qubit pulses, on the other hand, were up-mixed through a double-super-heterodyne \cite{horowitz_textbook_CUP_1989} setup employing two LOs and two single side-band mixers. These pulse generation setups also incorporated various amplifiers, filters, alternators, and fast microwave switches to achieve effective suppression of unwanted mixing products and to minimize leakage of LO signals. The signal from the refrigerator was down-mixed using the same readout LO and further amplified before being digitized by Operator X from Quantum Machines.

\newpage
\vspace{1cm}
\begin{center}
    \noindent\makebox[\textwidth]{%
      \includegraphics[width=183mm]{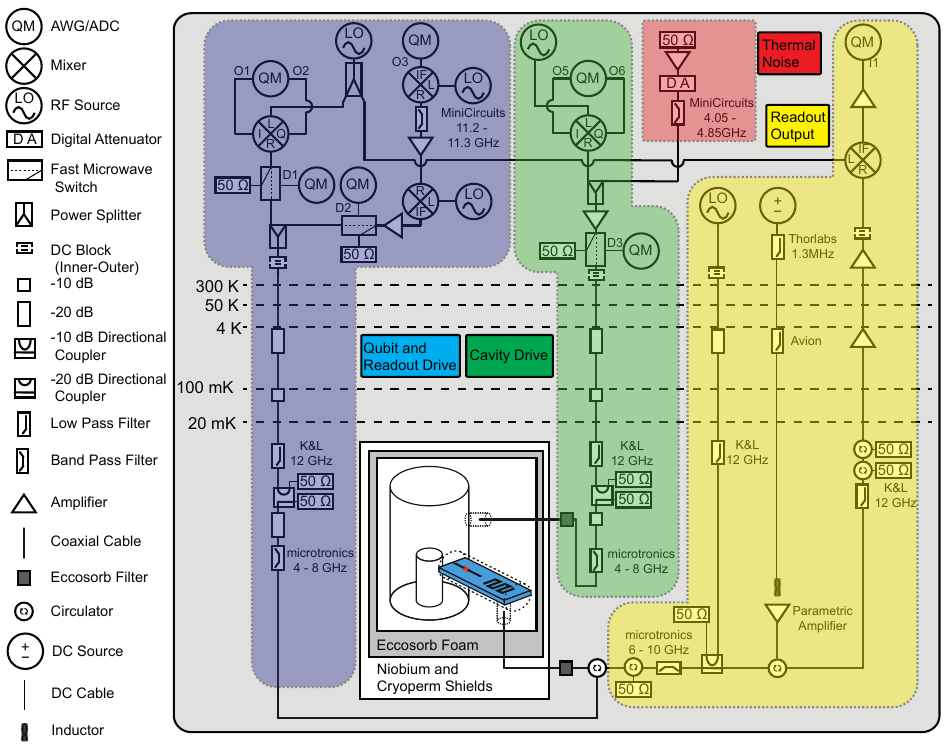}
    }
\end{center}
\noindent {\bf Fig. S1. Schematic of Experimental Wiring.} The cavity and readout resonator were driven by an IQ mixing setup while the qubit tones were up-converted via a double-super-heterodyne setup. The qubit-resonator line had a total of $70$ dB input line attenuation while the cavity had $60$ dB. Cavity noise was added via amplifying Johnson-Nyquist noise at room temperature by a total of $60$ dB and filtered with a MiniCircuits filter. The noise level was reduced by a digital attenuator. The readout tone was first amplified with a parametric amplifier before reaching the HEMTs. Noise from the output and amplifier pump lines were attenuated with isolators. Additional microwave and eccosorb filters were used to remove unwanted radiation from reaching the experiment. The setup was placed in superconducting and µ-metal shields and was surrounded by eccosorb foam.
\newpage

\subsection{\label{s:calibration}Calibration and Scaling of Wigner Function}
The Wigner function measurement was calibrated by the measurement of a single photon Fock state. The single photon Fock state was prepared by using a blue sideband transition. This technique is similar to that used in ion traps \cite{Diedrich_ion_sideband_cooling_1989} or atomic arrays \cite{Hamann_atom_sideband_cooling_1998}. The measurement data, which has arbitrary units, was collected in the pulse quadrature variables $I, Q$, which also have arbitrary units (Figure S2A). To calibrate the measurement, we seek a linear map from the data space into phase space. We find this map by fitting the function $\chi_W^{-1} W_{\ket 1}(\chi_I I + \ii\chi_{Q} Q)$ to the Fock state measurement data, where $\chi_W$, $\chi_I$, and $\chi_Q$ are scaling constants, and $W_{\ket 1}(\beta)$ is the Wigner function of the first Fock state (Figure S2B,C). Using the fitted scaling constants, we then map all other Wigner function measurement data into phase-space by constructing $\beta = \chi_I I + \ii\chi_{Q} Q$ and $W(\beta) = \chi_W D(I,Q)$, where $D(I,Q)$ is the measurement data.

\newpage
\begin{center}
    \noindent\makebox[\textwidth]{%
        \includegraphics[width=183mm]{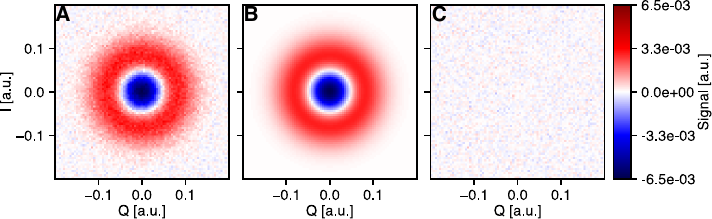}
    }
\end{center}
\noindent {\bf Fig. S2. Wigner Measurement and Fit of a Single Photon Fock State.} (\textbf{A}) Measured data and (\textbf{B}) analytical fit of a $\ket{1}$ cavity Fock state in data space. (\textbf{C}) Residuals of fit. We use the parameters of the fit to construct a linear map from data space into phase space (Section~\ref{s:calibration}).
\newpage

\subsection{\label{s:extra_measurements}Additional Hot Cat Wigner Function Measurements}
In Figure S3, we show additional Wigner function measurement maps done with the same protocol with an initial mean thermal cavity photon number of $n_{\mathrm{th}}=1.84(3)$.

In Figure S4, we report Wigner function measurements on states prepared by the qcMAP and ECD protocols in an earlier experimental setup. In this setup, we had $\chi_{\mathrm{qc}}/2\pi=1.272$ MHz, $K_\mathrm{c}/2\pi =2.33$ kHz, $\chi_{\mathrm{qc}}'/2\pi=7.1$ kHz, cavity lifetime $T_{1,\mathrm{\mathrm{c}}}= 128$ µs, qubit lifetime $T_{1}= 6.3 $ µs, and qubit coherence time $T_2^* = 2.4$ µs. In these experimental runs, the heat bath used to prepare the initial state was kept connected throughout the preparation and measurement protocol. Note that, in our setup, the cavity lifetime is limited by the external coupling via the coupling pin to the environment. Thus, the cavity coupling rate to the heat bath is the same as the cavity photon loss rate. The experiment was run with $\alpha=2.5$ and $n_\mathrm{th} = 2.07$.

\newpage
\begin{center}
    \includegraphics{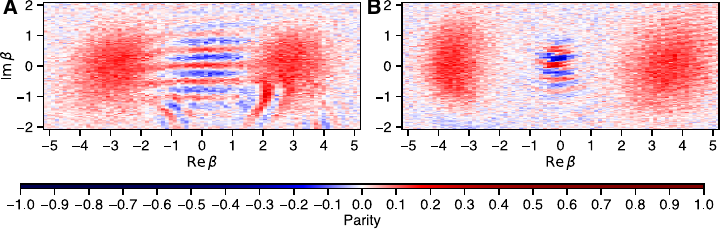}
\end{center}
\noindent{\bf Fig. S3. Additional hot cat state measurements.} (\textbf{A}) ECD protocol (\textbf{B}) qcMAP protocol. Note that this plot uses a linear scaling of the color bar. Here, the initial mean thermal cavity photon number is $n_{\mathrm{th}}=1.84(3)$.
\newpage

\newpage
\begin{center}
    \includegraphics{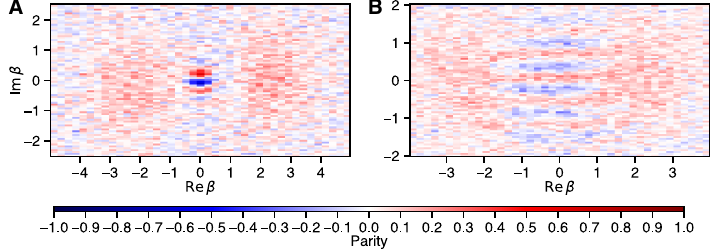}
\end{center}
\noindent {\bf Fig. S4. Hot cat state measurements from a previous experimental setup with the heat bath left connected.} (\textbf{A}) qcMAP protocol (\textbf{B}) ECD protocol. Note that this plot uses a linear scaling of the color bar. See section \ref{s:extra_measurements} for the parameters of this setup.
\newpage

\subsection{Characterizing the Initial Thermal State}
To determine the initial thermal state of the cavity mode, number-split qubit spectroscopy was performed (Figure S5). This allowed for the measurement of the cavity photon distribution. The steps involved are shown in Figure S5B. First, the cavity mode was equilibrated with the thermal bath as described in Section~\ref{s:setup}. Next, a cavity photon number selective qubit $\pi$-pulse was applied and the qubit state was measured. By repeating the measurement to get an ensemble average, the probability to excite the qubit at a certain frequency was determined.

Due to the cavity photon distribution and the dispersive coupling between the cavity and the qubit, the qubit resonance frequency is split into a spectrum of multiple frequencies where the relative resonance peak height depends on the cavity photon distribution (Figure S5C). By measuring the probability to excite the qubit across the frequency spectrum, we also directly measure the cavity photon number distribution. 

For a thermal state with average photon occupation number $n_\mathrm{th}$, the probability of measuring $n$ photons is
\begin{equation}
    P_{n_\mathrm{th}}(n) = \frac{n_\mathrm{th}^n}{(1+n_\mathrm{th})^{n+1}}.
    \label{Super-poissonian distribution}
\end{equation}
By fitting this to the spectral qubit excitation probability (Figure S5C), we determined $n_\mathrm{th}$ of the cavity state. By varying the added photon noise power, we found a relationship between the attenuator setting and $n_\mathrm{th}$ (Figure S5D).

The added thermal noise power is calculated from $P = k_B T f_\mathrm{BW}$. $f_\mathrm{BW}$ is the bandwidth of the cavity which is calculated from the cavity lifetime. $T$ is the temperature of the resistor which is at room temperature and $k_B$ is the Boltzman constant. The noise power is then reduced by the digital attenuator value. Critically, it is the measured mean thermal photon number or the cavity mode temperature that is used in the experiments and Fig. S5D just serves as a visual guide.

\newpage
\begin{center}
\includegraphics[width=121mm]{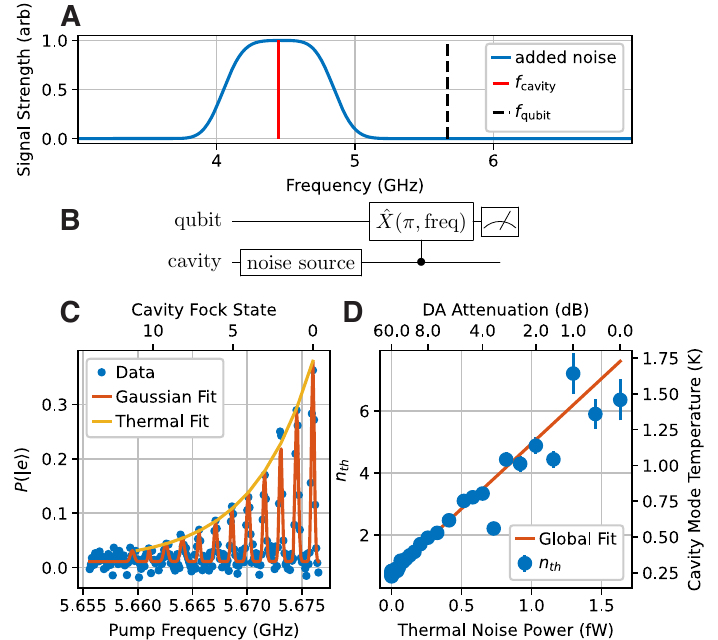}
\end{center}
\noindent {\bf Fig. S5. Thermal State Measurement Technique} (\textbf{A}) An illustration of frequency spectrum of the added noise. The thermal noise is only added at the cavity frequency while qubit frequencies are filtered out. The noise level is controlled by a digital attenuator. (\textbf{B}) Experimental Pulse sequence. (\textbf{C}) Qubit spectroscopy measurement result for a thermal state. Here, $n_{\mathrm{th}}=3.3(1)$. (\textbf{D}) Thermal population measurements for different attenuation settings,  corresponding to noise powers. The added noise power with the mean thermal photon is fitted with a straight line.
\newpage

\subsection{Characterization of Hamiltonian}
The full system Hamiltonian, including perturbative terms due to higher excitation levels of the qubit and cavity nonlinearities, is
\begin{equation}
\hat H/\hbar = \omega_{c} \hat c^{\dag} \hat c - \frac{K_{c}}{2} \hat c^{\dag} \hat c^{\dag} \hat c \hat c
+ \omega_{q} \hat q^{\dag} \hat q - \frac{K_{q}}{2} \hat q^{\dag} \hat q^{\dag} \hat q \hat q
- \chi_{qc} \hat c^{\dag} \hat c \hat q^{\dag} \hat q
- \frac{K'_{c}}{6} \hat c^{\dag} \hat c^{\dag} \hat c^{\dag} \hat c \hat c \hat c 
- \frac{\chi'_{qc}}{2} \hat c^{\dag} \hat c^{\dag} \hat c \hat c \hat q^{\dag} \hat q
\label{Hamiltonian}
\end{equation}
Here $\hat q^{\dag}$ and $\hat c^{\dag} $ are the creation operators for the qubit and cavity mode respectively, and $ \hat q,   \hat c$ are the corresponding annihilation operators. The values of the Hamiltonian parameters were measured experimentally and are reported in Table S1.

To characterize the Hamiltonian of our system, we employed a measurement method to accurately determine the cavity frequency as a function of the cavity photon number and qubit initial state (Figure S6A). A similar technique was reported in ~\cite{eickbusch_Nat.Phys.18:12_2022}. First, the cavity was displaced by $\beta$ and allowed to evolve for a time delay $t$, with the qubit in the ground state. The delay time was varied up to a maximum delay time, $T$. Subsequently, a second displacement with displacement parameter $\beta\exp\{-i \phi(t)\}$ was applied, where $\phi(t) = 2\pi \times 5 t / T$. Finally, a cavity ground state selective $\pi$-pulse (described by the operator $\hat{X}(\pi,\sigma_t)$ with $\sigma_t = 300$ ns) was applied to the qubit, and the qubit state was measured.

The principle of the measurement is illustrated in Figure S6B. For weak Kerr effects and the qubit in the ground state, the displaced state approximately evolves as $\ket{\beta \ee^{\ii\omega(\beta) t}}$, where 
\begin{equation}
    \omega(\beta) = \Delta - |\beta|^2 \frac{K_\mathrm{c}}{2} - |\beta|^4\frac{K_c'}{6},
    \label{omega_of_beta}
\end{equation} 
and the cavity frequency detuning $\Delta=\omega_{\mathrm{c}}-\omega_{\mathrm{drive}}$ from the drive frequency $\omega_{\mathrm{drive}}$. The ground-state selective $\pi$-pulse will flip the qubit only if $\phi(t)$ in the second displacement pulse matches $\omega(\beta)$, i.e.~$\phi(t)=\omega(\beta)$. Thus, the probability of measuring the cavity in the ground state, or equivalently, qubit in the excited state, is expected to be
\begin{equation}
    P(\ket e) = \lvert \braket{0|\beta(t)}\rvert ^2 = e^{-2|\beta|^2[1-\cos{(\omega(\beta)t)}] - t/T_{1,\mathrm{c}}},
    \label{Hamiltonian Measurement Result}
\end{equation}
The exponential decay comes from the finite cavity lifetime.

We measure $P(\ket e)$ as a function of $t$ and $\beta$ (Figure S6C). We then fit \eqnref{Hamiltonian Measurement Result} as a function of $t$ for the different values of $\beta$ used in the measurement (Figure S6D). From the fit, we extract $\omega(\beta)$ for the given value of $\beta$. We then fit \eqnref{omega_of_beta} to the measured $\omega(\beta)$ to extract $\Delta$, $K_c$, and $K_c'$ (Figure S6E). Finally, we repeat the procedure with the qubit initially in the excited state, which allows us to determine $\chi_\mathrm{qc}$ and $\chi_\mathrm{qc}'$.

\newpage
\begin{center}
    \noindent\makebox[\textwidth]{%
        \includegraphics[width=183mm]{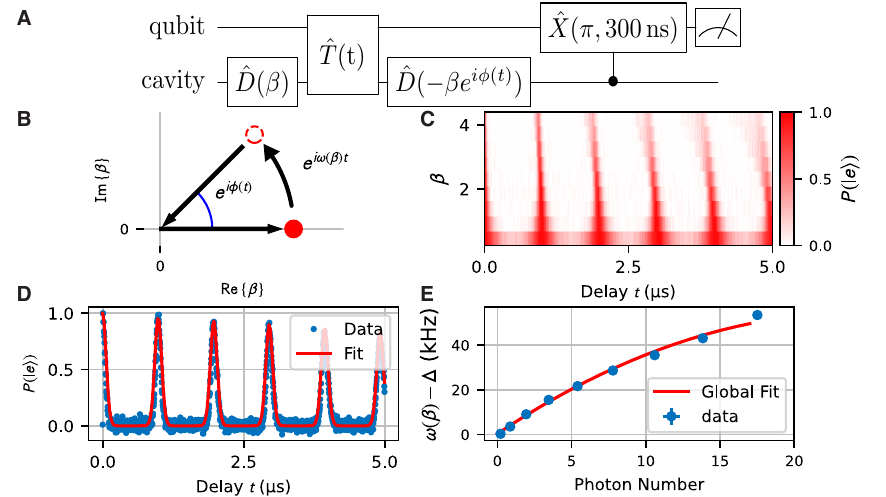}
    }
\end{center}
\noindent {\bf Fig. S6. Hamiltonian Measurement Technique} (\textbf{A}) Experimental pulse sequence. (\textbf{B}) Phase space evolution of the cavity during the experiment. (\textbf{C}) Measurement data of the qubit excited state probability for different values of the delay time and the initial displacement. Here, the qubit is initialized in the ground state. (\textbf{D}) Measurement of the qubit excited state probability for a fixed value of $\beta$. The solid line is a fit of \eqnref{Hamiltonian Measurement Result}.  (\textbf{E}) The photon-number dependent cavity frequency $\omega(\beta)$ as a function of the initial average photon number $|\beta|^2$ with a fit of \eqnref{omega_of_beta}.
\newpage

\newpage
\begin{center}
\noindent\makebox[\textwidth]{%
\begin{tabular}{||c |c | c ||} 
 \hline
 Parameter & Symbol & Value \\ 
 \hline\hline
 Qubit Frequency & $\omega_{\mathrm{q}}/2\pi$ & $5.676001$ GHz \\
 \hline
 Qubit Anharmonicity & $K_{\mathrm{q}}/2\pi$ & $189.9 \pm 0.4$ MHz \\
 \hline
 Qubit Lifetime & $T_{1}$ & $31.0 \pm 0.4$ µs \\
 \hline
 Qubit Coherence Time & $T_2^*$ & $12.5 \pm 0.4$ µs \\
 \hline
 Qubit Hann Echo Time & $T_2^E$ & $16.9 \pm 0.5$ µs \\ 
 \hline 
  High Q Cavity Frequency & $\omega_{\mathrm{c}}/2\pi$ & $4.544939$ GHz \\
 \hline
  High Q Cavity Self-Kerr & $K_{\mathrm{c}}/2\pi$ & $4.9 \pm 0.1$ kHz \\
 \hline
  High Q Cavity Second Order Self-Kerr & $K'_{\mathrm{c}}/2\pi$ & $14\pm 8$ Hz \\
 \hline
 High Q Cavity Lifetime & $T_{1,\mathrm{c}}$ & $110 \pm 2$ µs \\
 \hline
 High Q Cavity - Qubit Dispersive Shift & $\chi_{\mathrm{qc}}/2\pi$ & $1.499 \pm 0.003$ MHz \\
 \hline
 High Q Cavity - Qubit Second Order Dispersive shift &  $\chi'_{\mathrm{qc}}/2\pi$ & $12.8 \pm 0.9$ kHz \\
 \hline
 High Q Cavity Residual Mean Thermal Photon Number & $n_{\mathrm{th,\ residual}}$ & $0.0338 \pm 0.007$ \\
 \hline
 High Q Cavity Residual Mode Temperature & $T_{\mathrm{cavity,\ residual}}$ & $63.7 \pm 0.4$ mK \\
 \hline
 Qubit Residual Mode Temperature & $T_{\mathrm{qubit}}$ & $68 \pm 1$ mK \\ 
 \hline 
  Readout Resonator Frequency & $\omega_{\mathrm{r}}/2\pi$ & $7.528852$ GHz \\
 \hline
  Readout Resonator External Coupling & $\kappa_{\mathrm{c\ ext}}/2\pi$ & $1.33$ MHz \\
 \hline
 Readout Resonator - Qubit Dispersive Shift & $\chi_{\mathrm{rq}}/2\pi$ & $1.61$ MHz \\
 \hline
 Readout Resonator - High Q Cavity Dispersive Shift (calculated) & $\chi_{\mathrm{rc, cal}}/2\pi$ & $3.4$ kHz \\
 \hline  
\end{tabular}
}
\end{center}
\noindent {\bf Table. S1. Hamiltonian Parameters. } The high Q cavity lifetime is limited by its external coupling. The reported cross Kerr $\chi_{\mathrm{readout - cavity, cal}}$ is a lower bound calculated via  $\chi_{\mathrm{rc}}= \chi_{\mathrm{qc}}\chi_{\mathrm{qr}} (\frac{1}{\Delta_{\mathrm{qc}}}+\frac{1}{\Delta_{\mathrm{qr}}})$.
\newpage

\subsection{Density matrix reconstruction}
\noindent In general, the density matrix $\hat\rho$ can be obtained from the Wigner function $W(\beta)$ as 
\begin{equation}
    \hat\rho = 2\int \dd^2\beta\, W(\beta) \hat\varPi(\beta).
    \label{dm_from_wigner}
\end{equation}
This equation is the inverse of $W(\beta) = \frac{2}{\pi}\tr\{\hat \rho \hat \varPi(\beta)\}$ and can be shown by writing out $\hat\varPi(\beta)$ in a known basis, e.g.~$\hat\varPi(\beta) = \int_{-\infty}^\infty \dd x\, \ee^{-2\ii x \sqrt 2\Im\{\beta\}}\ket{\sqrt 2\Re\{\beta\} - x}\bra{\sqrt 2\Re\{\beta\} + x}$ where $\ket x$ is an eigenket of $\hat x = (\hat c + \hat c^\dagger)/\sqrt 2$ \cite{royer_Phys.Rev.A_15:2_1977}. To reconstruct the density matrices of the states prepared, we choose to compute the matrix elements in the Fock basis
\begin{equation}
    \rho_{mn} = \bra{m} \hat\rho \ket {n} = 2\int\dd^2\beta\, W(\beta) \bra{m}\hat\varPi(\beta)\ket{n}
    \label{fock_dm_from_wf}
\end{equation}
using our measured data for $W(\beta)$. The operator $\hat\varPi(\beta)$ has Fock basis matrix elements~\cite{barnett_MethodsTheoretical_2002}
\begin{equation}
        \bra m \hat\varPi(\beta) \ket n = (-1)^n \cdot \left\{\begin{aligned}& \sqrt{\frac{n!}{m!}} (2\beta)^{m-n} \ee^{-2|\beta|^2} \mathcal{L}^{(m-n)}_n(4|\beta|^2) & (m \geq n) \\
        & \sqrt{\frac{m!}{n!}} (-2\beta^*)^{n-m} \ee^{-2|\beta|^2} \mathcal{L}^{(n-m)}_m(4|\beta|^2) & (n\geq m)
    \end{aligned}
    \right.
\end{equation}
where $\mathcal{L}^{(k)}_n(x)$ are the associated Laguerre polynomials. The Wigner function data is given as averages on a grid $\beta_{rj}$ corresponding to experimentally measured values of $\beta$, with $N_R,N_I$ the dimensions of the grid and $\Delta_{R,I}$ the grid spacing along the real and imaginary directions. The measured Wigner function data is the $N_R \times N_I$ matrix $W_{rj} = W(\beta_{rj})$. We can then discretize \eqnref{fock_dm_from_wf} on $\beta_{rj}$ to find
\begin{equation}
    \rho_{mn} \approx 2 \Delta_R\Delta_I \sum_{r=0}^{N_R-1}\sum_{j=0}^{N_I-1} W_{rj} \varPi_{mn,rj}
    \label{dm_elements_estimate}
\end{equation}
which is approximate due to the finite size of the grid, and where we defined
\begin{equation}
    \varPi_{mn,rj} = \bra m \hat\varPi(\beta_{rj}) \ket n.
\end{equation}
With this method, we can estimate the density matrix directly from the data. Recall that a proper density matrix must have unit trace $\tr{\hat\rho} = 1$, be hermitian $\hat\rho^\dagger = \hat\rho$, and be positive semidefinite. The estimate $\rho_{mn}$ obtained from \eqnref{dm_elements_estimate} will be hermitian, but not strictly normalized nor positive semidefinite. This is for three reasons: 1) The measurements contain additive randomly distributed noise which propagates into the matrix $\rho_{mn}$. 2) The data grid has a finite size, which makes \eqnref{dm_elements_estimate} be an inexact approximation to \eqnref{dm_from_wigner}, in particular for large $m$ and $n$. 3) As explained in Section~\ref{s:measurement_theory_model}, the Wigner measurement does not strictly produce a Wigner function but rather the function
\begin{equation}
    W_\mathrm{meas.}(\beta) = p_g W_{g}(\beta) - p_e W_{e}(\beta)
\end{equation}
where $p_{g,e}$ are the qubit $\ket g$ and $\ket e$ populations at the end of the state preparation, and $W_{g,e}(\beta)$ are respectively the Wigner functions corresponding to the reduced density matrices $\hat\rho_{g} = \bra g \hat\rho_\mathrm{tot} \ket g$ and $\hat\rho_{e} = \bra e \hat\rho_\mathrm{tot} \ket e$ of the total oscillator-qubit density matrix $\hat\rho_\mathrm{tot.}$. Due to the linearity of \eqnref{dm_from_wigner}, our density matrix reconstruction therefore gives us the matrix elements
\begin{equation}
    \mu_{mn} = \bra m p_g \hat \rho_{g} - p_e \hat\rho_{e} \ket n.
\end{equation}
As $\hat\rho_{e}$ is positive semidefinite and $p_{e} \geq 0$, the term $-p_e \hat\rho_{e}$ is negative semidefinite and thus makes the matrix $\mu$ indefinite under the conditions of our experiment. Out of the three error sources, we find 3) to be the most important in our case. We are seeking to estimate the density matrix $\hat\rho_{g}$, with $\hat\rho_{e}$ a perturbation. Efficient max-likelihood estimators for the density matrix such as~\cite{smolin_Phys.Rev.Lett._108:070502_2012} are not designed for this scenario. We therefore choose to obtain our final estimate for $\rho_{mn}$ by truncating all negative eigenvalues of $\mu$. More specifically, we diagonalize $\mu = U D U^\dagger$, where $D$ is the diagonal matrix of eigenvalues and $U$ the matrix of eigenvectors of $\mu$, and then construct the matrix $\rho = U \max(D, 0) U^\dagger /\tr\{\max(D, 0)\}$, where the $\max$ operation applies elementwise. This provides our final estimate $\rho_{mn}$, which we plot in Figure 1H-J in the main text. 

Using $\rho_{mn}$ as the reconstructed density matrix of the prepared state, we can compute its representation in the eigenbasis of the position operator $\hat x = (\hat a + \hat a^\dagger)/\sqrt 2$, and then its coherence function.
The position representation density matrix can be computed as
\begin{equation}
    \rho(x_1,x_2) = \sum_l p_l [U\vb \psi(x_1)]_l [U \vb \psi(x_2)]_l^*
\end{equation}
where $p_l$ are the eigenvalues of $\rho_{mn}$, $\vb \psi(x)$ is a vector with elements $\bra x n \rangle = \ee^{-x^2/2} H_n(x) / \sqrt{2^n n! \sqrt\pi}$ where $H_n(x)$ are the Hermite polynomials, and $U$ is the eigenvector matrix of $\rho_{mn}$. Using this expression for $\rho(x_1,x_2)$, we compute the coherence function as
\begin{equation}
    g(x_1,x_2) = \frac{|\rho(x_1,x_2)|}{\sqrt{\rho(x_1,x_1)\rho(x_2,x_2)}}.
\end{equation}
 Using this method, we obtain the coherence function estimates plotted in Figure 4 in the main text.

\newpage
\subsection{\label{s:parity_lifetime}Hot cat state fringe lifetime measurement}

We measure the decay of the fringes of the hot cat states by measuring the parity of the cat state at the phase space location (0,0) after a delay time.

The fringe lifetime is related to the cavity lifetime by \cite{kim_Phys.Rev.A_46:7_1992}
\begin{equation}
    T_{\mathrm{fringe}} = \frac{T_{\mathrm{1,c}}}{2 |\alpha|^2 (2 n_\mathrm{b}+1)},
\end{equation}
where $\alpha$ is the cat size and $n_\mathrm{b}$ is the mean thermal excitation number of the bath the system is connected to (note that this is not the initial mean thermal photon population). Using the values in Table S1, we estimate the expected fringe lifetime to be $4.7\pm 0.9$ µs.

Figure S7 shows the measurement of the fringe lifetime for hot cat states. The average lifetime across all hot cat states is $ 3.7 \pm 0.2$ µs. The discrepancy with the expected lifetime indicates the presence of decoherence channels beyond those characterised in Table~S1.

\newpage
\begin{center}
    \noindent\makebox[\textwidth]{%
        \includegraphics{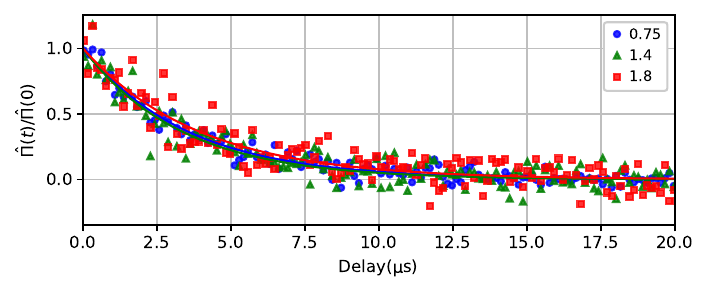}
    }
\end{center}
\noindent {{\bf Fig. S7. Hot cat state fringe lifetime.} The label refers to the initial thermal excitation number in the cavity $n_\mathrm{th}$. The y-values are normalized to the initial parity of the fringe at zero delay. The lines are exponential fits from which the average lifetime across all hot cat states is obtained to be $ 3.7 \pm 0.2$ µs.}

\newpage
\section{\label{s:theory}Theory}

We show that the qcMAP and ECD protocols are described by the operators $\hat S_{1,2}$ under ideal conditions, derive the hot cat Wigner functions $W_{1,2}(\beta)$, and analyze the coherence functions of the hot cats.

\subsection{\label{s:pulse_sequence_analysis}Theoretical analysis of the qcMAP and ECD protocols}

In this section, we show that the ECD and qcMAP pulse sequences displayed in the quantum circuit diagrams in Figure 2A-C are respectively equivalent to applying the operators $\hat S_1'$ and $\hat S_2$ to the initial cavity state $\hat\rho_0$. A necessary condition is that $\hat\rho_0$ has no overlap with itself when displaced by $2\alpha$. We repeat the operator definitions from the main text for convenience. The equivalent operator for the ECD protocol is
\begin{equation}
    \hat S_1' \equiv \frac{1}{\sqrt 2}\left[\hat D(\alpha) - \ee^{\ii(\phi + 2|\alpha|^2)}\hat D(-\alpha)\right]\ii^{\hat n},
    \label{S1'_def}
\end{equation}
which in turn is equivalent to the operator
\begin{equation}
    \hat S_1 \equiv \frac{1}{\sqrt 2}\left[\hat D(\alpha) + \ee^{\ii\phi}\hat D(-\alpha)\right]
\end{equation}
when the initial state is thermal and $\phi$ is experimentally controllable.
The equivalent operator for the qcMAP protocol is
\begin{equation}
    \hat S_2 \equiv \frac{1}{\sqrt 2}\left[1 - \ee^{\ii\phi}\hat \varPi\right]\hat D(\alpha).
\end{equation}

From Figure 2A-C, we read that the ECD and qcMAP protocols first prepare the initial thermal state $\hat\rho_0$ of the cavity and then apply the unitary operators
\begin{equation}
    \hat U_1 \equiv \hat D(\alpha)\hat X(\pi,\sigma_t) \hat D(-\alpha)\hat T(\pi/2\chi_\mathrm{qc}) \hat D[-\alpha\tfrac{(1+\ii)}{2}]\hat Y(\pi)\hat D[-\alpha\tfrac{(1+\ii)}{2}] \hat T(\pi/2\chi_\mathrm{qc})\hat D(\alpha)\hat X(\pi/2,\phi)
    \label{echo_pulse_sequence_op}
\end{equation}
for ECD, and
\begin{equation}
    \hat U_2 \equiv \hat D(-\alpha)\hat X(\pi,\sigma_t)\hat D(\alpha)\hat T(\pi/\chi_\mathrm{qc})\hat D(\alpha)\hat X(\pi/2,\phi)
    \label{qcmap_pulse_sequence_op}
\end{equation}
for qcMAP, 
to the cavity-qubit initial state $\hat\rho_0 \ket g \bra g$ (we define all operators appearing in these expressions in the next paragraph). Under ideal conditions, the final state of the protocols is 
\begin{equation}
    \hat U_i \hat \rho_0 \ket g \bra g \hat U_i^\dagger = \hat\rho_i \ket g \bra g
    \label{final_state}
\end{equation}
($i \in \{1,2\}$), with
\begin{equation}
    \hat\rho_i \equiv \hat S_{gg,i} \hat\rho_0 \hat S_{gg,i}^\dagger,
\end{equation}
\begin{equation}
    \hat S_{gg,i} \equiv \bra g \hat U_i \ket g.
    \label{equiv_op_definition}
\end{equation}
It is straightforward to show that the final cavity-qubit state \eqnref{final_state} is a product state, with the qubit in the $\ket g$ state, if and only if
\begin{equation}
    \tr\{\hat S_{gg,i} \hat\rho_0 \hat S_{gg,i}^\dagger\} = 1.
    \label{unitarity_condition}
\end{equation}
Consequently, the ECD and qcMAP protocols, given exactly by $\hat U_{1,2}$, are equivalent to the operators $\hat S_{gg,\{1,2\}}$ for all initial states $\hat\rho_0$ which satisfy \eqnref{unitarity_condition}.

The operators appearing in Eqs.~(\ref{echo_pulse_sequence_op}) and (\ref{qcmap_pulse_sequence_op}) are: 
I) The cavity displacement operator
\begin{equation}
    \hat D(\beta) \equiv \exp{\left\{\beta \hat c^\dagger - \beta^*\hat c\right\}}
\end{equation}
where $\hat c,\hat c^\dagger$ are the cavity annihilation and creation operators and $\beta$ is a complex-valued argument. II) The time-evolution operator under the dispersive interaction Hamiltonian $\hat T(t)$. The interaction Hamiltonian is $\hat H = -\hbar\chi_{\mathrm{qc}}\hat n\ket e \bra e$, where $\hat n = \hat c^\dagger \hat c$, so the time-evolution operator is
\begin{equation}
    \hat T(t) = \exp{\left\{\ii t \chi_\mathrm{qc} \hat n \ket e \bra e \right\}} = \ket g \bra g + \exp{\left\{\ii t \chi_\mathrm{qc} \hat n \right\}}\ket e \bra e.
\end{equation}
In particular,
\begin{equation}
    \hat T(\pi/\chi_\mathrm{qc}) = \ket g \bra g + \hat \varPi\ket e \bra e
\end{equation}
where $\hat\varPi = (-1)^{\hat n}$ is the parity operator, and
\begin{equation}
    \hat T(\pi/2\chi_\mathrm{qc}) = \ket g \bra g + \ii^{\hat n}\ket e \bra e.
\end{equation}
In phase space, the operator $\ii^{\hat n}$ leads to a counterclockwise rotation by $\pi/2$ around the phase-space origin.
III) The qubit rotation operators acting only on the qubit Hilbert space
\begin{equation}
    \hat X(\pi/2,\phi) \equiv \frac{1}{\sqrt 2}\left[\left(\ket g + \ii\ee^{\ii\phi}\ket e\right)\bra g + \left(\ket e + \ii\ee^{-\ii\phi}\ket g\right)\bra e\right],
\end{equation}
where $\phi$ is an experimentally controllable phase,
\begin{equation}
    \hat X(\pi) \equiv \ii\left(\ket g \bra e + \ket e \bra g\right),
\end{equation}
and
\begin{equation}
    \hat Y(\pi) \equiv \hat R(\pi,\pi/2) = \ket g \bra e - \ket e \bra g.
\end{equation}
These operators are special cases of the general qubit rotation operator
\begin{equation}
    \hat R(a,\vb u) \equiv \exp\left\{\frac{\ii a}{2} \vb u \cdot \hat{\vb \sigma} \right\} = \hat 1 \cos\frac{a}{2} + \ii \vb u \cdot \hat{\vb \sigma} \sin\frac{a}{2}
    \label{su2_euler}
\end{equation}
with $a$ a real variable, $\vb u$ a unit vector on the 3-dimensional unit sphere, and $\hat{\vb \sigma} = \hat\sigma_x \vb e_x + \hat \sigma_y \vb e_y + \hat \sigma_z \vb e_z$ ($\vb e_j$ are the coordinate unit vectors and $\hat \sigma_j$ the Pauli matrices $\hat \sigma_x = \ket g \bra e + \ket e \bra g$, $\hat \sigma_y = \ii (-\ket g \bra e + \ket e \bra g)$, $\hat \sigma_z = \ket g \bra g - \ket e \bra e$). We have $\hat X(\pi/2,\phi) = \hat R(\pi/2, \cos\phi\,\vb e_x + \sin\phi\,\vb e_y)$, $\hat X(\pi) = \hat R(\pi,\vb e_x)$, $\hat Y(\pi) = \hat R(\pi, \vb e_y)$. IV) The cavity-selective qubit rotation operator $\hat X(\pi, \sigma_t)$. We assume that it takes the form
\begin{equation}
    \hat X(\pi, \sigma_t) = \sum_{n=0}^\infty \ket n \bra n \hat R(a_n, \vb u_n),
    \label{XpiN_def}
\end{equation}
i.e.~that it is defined by specifying a sequence of qubit rotation operators on the Fock states $\ket n$. We further assume that the parameters $\{a_n\}_{n=0}^\infty$ and $\{\vb u_n\}_{n=0}^\infty$ are such that for $n \leq N$, where $N$ is a number that depends on $\sigma_t$, $a_n = \pi$ and $\vb u_n = \vb e_x$. Additionally, the sequence of parameters $\{a_n\}_{n=0}^\infty$ decays so that there is a number $M$ (also determined by $\sigma_t$) such that $a_n = 0$ when $n > M$. In this case, we can write
\begin{equation}
    \hat X(\pi, \sigma_t) = \ii \hat P_{\leq N} \hat\sigma_x  + \hat P_{>M} + \hat Q_{NM},
    \label{disentanglement_pulse_operator_decomposition}
\end{equation}
where we have introduced the operators
\begin{equation}
    \hat P_{\leq N} \equiv \sum_{n=0}^N \ket n \bra n,
\end{equation}
\begin{equation}
    \hat P_{> M} \equiv \sum_{n=M+1}^\infty \ket n \bra n,
\end{equation}
\begin{equation}
    \hat Q_{NM} \equiv \sum_{n=N+1}^M \ket n \bra n \hat R(a_n, \vb u_n).
\end{equation}
It is possible (by using a Magnus approximation \cite{thomas_Phys.Rev.A_27:5_1983}) to find an explicit expression for the operator $\hat X(\pi, \sigma_t)$ that results when a Gaussian qubit pulse is applied to our dispersively coupled cavity-qubit system. That expression is indeed well-described by \eqnref{disentanglement_pulse_operator_decomposition}. However, for the present discussion, we do not need to specify $\hat X(\pi, \sigma_t)$ beyond the description already made.

We now show the equivalence of $\hat U_{1,2}$ to $\hat S_{1,2}$ under conditions which we identify during the analysis. We begin with the qcMAP protocol $\hat U_2$. From the definitions made, $\hat U_2$ can be rewritten as
\begin{equation}
    \hat U_2 = \frac{1}{\sqrt 2}\hat D(-\alpha)\hat X(\pi, \sigma_t)\left[\hat D(2\alpha) \ket g \left(\bra g + \ii\ee^{\ii\phi}\bra e\right) + \hat\varPi \ket e \left(\ii\ee^{\ii\phi}\bra g + \bra e\right) \right]
    \label{u2}
\end{equation}
from which we identify
\begin{multline}
    \hat S_{gg,2} = \frac{1}{\sqrt 2}\hat D(-\alpha)\left[\bra g \hat X(\pi, \sigma_t) \ket g\hat D(2\alpha) + \ii\ee^{\ii\phi} \hat\varPi \bra g\hat X(\pi, \sigma_t) \ket e\right] \\
    = \frac{1}{\sqrt 2}\hat D(-\alpha)\left[\left(\hat P_{>M} + \bra g \hat Q_{NM} \ket g\right)\hat D(2\alpha) + \ii\ee^{\ii\phi} \hat\varPi \left(\ii\hat P_{\leq N} + \bra g \hat Q_{NM} \ket e\right)\right]
    \label{Sgg2_def}
\end{multline}
where we inserted \eqnref{disentanglement_pulse_operator_decomposition} to go from the first to the second line. We now wish to identify the initial states $\hat\rho_0$ for which $\hat S_{gg,2}$ fulfils \eqnref{unitarity_condition}. Consider first an initial state $\hat\rho_0$ which has non-zero matrix elements only in the first $N$ Fock states, i.e.
\begin{equation}
    \hat\rho_0 = \sum_{k,l=0}^N \bra k \hat\rho_0 \ket l \ket k \bra l.
    \label{finite_Fock_initial_state}
\end{equation}
For this $\hat\rho_0$, $\hat P_{\leq N}\hat\rho_0 = \hat\rho_0$. If we have additionally chosen $\alpha$ large enough so that the matrix element $\bra M \hat D(2\alpha) \ket M$ is negligible, then it also follows (using the triangle inequality) that $\hat P_{\leq M}\hat D(2\alpha)\hat\rho_0 = 0$. In a Wigner function picture, the condition on $\alpha$ can be formulated as the $M$:th Fock state Wigner function being negligible for arguments larger than $|\alpha|$, i.e.~$W_\ket M(|\beta| \geq |\alpha|) = 0$. From $\hat P_{\leq N}\hat\rho_0 = \hat\rho_0$ and $\hat P_{\leq M}\hat D(2\alpha)\hat\rho_0 = 0$ it follows that $\hat Q_{NM}\hat\rho_0 = 0$, $\hat Q_{NM}\hat D(2\alpha)\hat\rho_0 = 0$, and $\hat P_{>M}\hat D(2\alpha)\hat\rho_0 = (1 - \hat P_{\leq M})\hat D(2\alpha)\hat\rho_0 = \hat D(2\alpha)\hat\rho_0$. We therefore obtain
\begin{equation}
    \hat S_{gg,2}\hat\rho_0 = \frac{1}{\sqrt 2}\left[\hat D(\alpha) - \ee^{\ii\phi}\hat D(-\alpha)\hat \varPi\right]\hat\rho_0 = \hat S_{2}\hat\rho_0.
    \label{Sgg2_S2_eq1}
\end{equation}
The condition \eqnref{unitarity_condition} is also satisfied. Consequently, $\hat U_2\hat\rho_0 \ket g \bra g \hat U_2^\dagger = \hat S_2\hat\rho_0\hat S_2^\dagger \ket g \bra g$, and $\hat S_2$ accurately describes the action of the qcMAP protocol, for all $\hat\rho_0$ of the form \eqnref{finite_Fock_initial_state} when $\alpha$ is such that $W_\ket{M}(|\beta| \geq |\alpha|) = 0$. 

Identical arguments show the equivalence of $\hat U_1$ to $\hat S_1'$ . From \eqnref{echo_pulse_sequence_op}, $\hat U_1$ can be written
\begin{equation}
    \hat U_1 = \frac{1}{\sqrt 2}\hat D(\alpha)\hat X(\pi, \sigma_t)\left[\hat D(-2\alpha)\ee^{\ii|\alpha|^2}\ket g\left(\bra g \ii \ee^{\ii\phi} + \bra e \right) - \ee^{-\ii|\alpha|^2}\ket e\left(\bra g + \ii\ee^{-\ii\phi}\bra e\right)\right]\ii^{\hat n}.
    \label{u1}
\end{equation}
We identify
\begin{equation}
    \hat S_{gg,1} = \frac{1}{\sqrt 2}\hat D(\alpha)\left[\ii \ee^{\ii\phi}\left(\hat P_{>M} + \bra g \hat Q_{NM}\ket g\right)\hat D(-2\alpha)\ee^{\ii|\alpha|^2} - \ee^{-\ii|\alpha|^2}\left(\ii \hat P_{\leq N} + \bra g \hat Q_{NM} \ket e\right)\right]\ii^{\hat n}.
    \label{Sgg1_def}
\end{equation}
When this operator acts on a state $\hat\rho_0$ described by \eqnref{finite_Fock_initial_state}, and $\alpha$ is large enough so that $\hat P_{\leq M}\hat D(2\alpha)\hat\rho_0 = 0$ as before, then
\begin{equation}
    \hat S_{gg,1}\hat\rho_0 = \frac{-\ii\ee^{-\ii|\alpha|^2}}{\sqrt 2}\left[\hat D(\alpha) - \ee^{\ii(\phi + 2|\alpha|^2)}\hat D(-\alpha)\right]\ii^{\hat n}\hat\rho_0 = -\ii\ee^{-\ii|\alpha|^2}\hat S_1'\hat\rho_0.
\end{equation}
\eqnref{unitarity_condition} is satisfied. The global phase vanishes when considering $\hat S_{gg,1}\hat\rho_0\hat S_{gg,1}^\dagger$.

We now relax the assumption that $\hat\rho_0$ is of the form \eqnref{finite_Fock_initial_state}. In general, the state after the ECD and qcMAP protocols can be written
\begin{equation}
    \hat U_i \hat \rho_0 \ket g \bra g \hat U_i^\dagger = p_g \hat\rho_g \ket g \bra g + p_e \hat\rho_e \ket e \bra e + \hat\psi
    \label{total_cavit_qubit_state}
\end{equation}
where
\begin{equation}
    p_g = \tr\{\hat S_{gg,i}\hat\rho_0\hat S_{gg,i}^\dagger\},
    \label{pg_def}
\end{equation}
\begin{equation}
    p_e = 1 - p_g,
\end{equation}
are the probabilities of finding the qubit in the $\ket g$ and $\ket e$ states,
\begin{equation}
    \hat \rho_g \equiv \frac{\hat S_{gg, i} \hat\rho_0 \hat S_{gg,i}^\dagger}{p_g},
\end{equation}
\begin{equation}
    \hat \rho_e \equiv \frac{\hat S_{eg, i} \hat\rho_0 \hat S_{eg,i}^\dagger}{p_e}
\end{equation}
are the qubit-conditional cavity states (where $\hat S_{eg,i} \equiv \bra e \hat U_{i} \ket g$), and $\hat\psi$ represents off-diagonal terms in the qubit basis. The previous condition \eqnref{unitarity_condition} is $p_g = 1$. For initial states $\hat\rho_0$ which have non-negligible matrix elements only for Fock numbers $\leq N$, i.e.~are described by \eqnref{finite_Fock_initial_state} plus a negligible part, one has $p_g \approx 1$, $p_e \approx 0$. More precisely, one computes
\begin{equation}
    p_g = \frac{1}{2}\left[\langle \hat P_{\leq N} \rangle + \langle \hat D(-2\alpha)\hat P_{>M} \hat D(2\alpha) \rangle + q(\sigma_t,\alpha,n_\mathrm{th})\right]
\end{equation}
where the expectation value is with respect to $\hat \rho_0$, and $q$ represents terms related to expectation values of $\hat Q_{NM}$. If both the first two terms are $1$, then $q(\sigma_t,\alpha,n_\mathrm{th}) = 0$. If $\langle \hat P_{\leq N} \rangle \approx 1$, it is always possible to choose $\alpha$ such that also $\langle \hat D(-2\alpha)\hat P_{>M} \hat D(2\alpha) \rangle \approx 1$ and therefore $q(\sigma_t,\alpha,n_\mathrm{th}) \approx 0$. For these states, the equivalence of $\hat U_{1,2}$ to $\hat S_{1,2}$ can therefore be satisfied to arbitrary precision in principle, and it is a good approximation to consider the ECD and qcMAP protocols equivalent to $\hat S_{1,2}$.
The thermal state
\begin{equation}
    \hat\rho_T \equiv \frac{1}{n_\mathrm{th} + 1}\sum_{n=0}^\infty \left(\frac{n_\mathrm{th}}{n_\mathrm{th} + 1}\right)^n\ket n \bra n,
    \label{thermal_state_def}
\end{equation}
is a particular example of a state which can be considered to have negligible representation for Fock numbers above $N$. It has 
\begin{equation}
    \tr\{\hat P_{> N}\hat\rho_T\} = \left(\frac{n_\mathrm{th}}{n_\mathrm{th} + 1}\right)^{N+1} \label{thermal_disentangling_projection_condition}
\end{equation}
which goes to $0$ in the limit $N\to\infty$.
In practice, the choice of $\sigma_t$ (i.e.~$N$ and $M$) and $\alpha$ are restricted by experimental limitations, and the nonzero $p_e$ leads to perturbations from the ideal result (i.e.~that described by $\hat S_{1,2}$).

The conditions derived so far are those under which our protocols become equivalent to the operators $\hat S_{1,2}$. As a final remark, we note that one can also study the conditions for the operators $\hat S_{1,2}$ to be effectively unitary independently of how they are implemented. This can be done by replacing $\hat S_{gg,i}$ with $\hat S_i$ in \eqnref{unitarity_condition}. For a general initial state $\hat\rho_0$, one then has
\begin{equation}
    \tr\left\{\hat S_{1}\hat\rho_0\hat S_{1}^\dagger\right\} = 1 - \Re\left\{\ee^{-\ii\phi}\chi_0(2\alpha)\right\} = 1,
    \label{S1_unitarity_condition}
\end{equation}
\begin{equation}
    \tr\left\{\hat S_{2}\hat\rho_0\hat S_{2}^\dagger\right\} = 1 - \frac{\pi}{2}\cos{\phi}\, W_0(\alpha) = 1.
    \label{S2_unitarity_condition}
\end{equation}
Here, $W_0(\alpha) = 2\pi^{-1}\tr\{\hat\varPi(\alpha)\hat\rho_0\}$ and $\chi_0(\alpha) = \tr\{\hat D(\alpha)\hat\rho_0\}$ are the Wigner and characteristic functions of the initial state. These equations can be satisfied either by choice of $\phi$ or $\alpha$. If we want this equation to be satisfied for any $\phi$, this gives the necessary conditions that $\chi_0(|\beta| > 2|\alpha|) = 0$ ($\hat S_1$) and $W_0(|\beta| > |\alpha|) = 0$ ($\hat S_2$). We also see that if $\phi$ is a half-integer multiple of $\pi$, the operator $\hat S_2$ is unitary independently of $\alpha$. For states that have a constant-phase characteristic function (e.g. thermal states), there are also choices of $\phi$ for which $\hat S_{1}$ is effectively unitary independently of $\alpha$. To avoid confusion, we stress that $\phi$ does not enter any of the conditions for the ECD and qcaMAP protocols to be equivalent to the operators $\hat S_{1,2}$. 

\subsection{\label{s:general_wigner_functions}Derivation of the Wigner functions $W_{1,2}(\beta)$ from the operators $\hat S_{1,2}$}

We give derivations of the Wigner functions $W_{1,2}(\beta)$ that result when the operators $\hat S_{1,2}$ are applied to an initial state $\hat\rho_0$ and in particular the thermal state $\hat\rho_T$. Starting from
\begin{equation}
    W_{1,2}(\beta) = \frac 2 \pi \tr\{\hat \varPi(\beta)\hat S_{1,2}\hat\rho_0\hat S_{1,2}^\dagger\}
\end{equation}
where $\hat\varPi(\beta) = \hat D(\beta)\hat \varPi \hat D^\dagger(\beta)$, one uses the cyclic property of the trace to express $W_{1,2}(\beta)$ as the expectation value of the operator $\hat S_{1,2}^\dagger\hat \varPi(\beta)\hat S_{1,2}$ in the initial state. This allows us to express the Wigner functions $W_{1,2}(\beta)$ in terms of the Wigner function of the initial state $W_0(\beta)$, which is given by
\begin{equation}
    W_0(\beta) = \frac{2}{\pi}\tr\{\hat \varPi(\beta)\hat\rho_0\}.
\end{equation}
Here, $\hat\rho_0$ is any state for which the operators $\hat S_{1,2}$ accurately describe the outcome of the ECD and qcMAP protocols (see Section~\ref{s:pulse_sequence_analysis}). One computes
\begin{equation}
    \hat S^\dagger_1\hat \varPi(\beta) \hat S_1 = \frac{1}{2}\Bigl\{\hat\varPi\left(\beta-\alpha\right) + \hat\varPi\left(\beta + \alpha\right) \\ - 2\cos\left(4\Im\{\alpha^*\beta\} + \phi\right) \hat\varPi(\beta) \Bigr\}
\end{equation}
and consequently
\begin{equation}
    W_1(\beta) = \frac{1}{2}\Bigl\{ W_0\left(\beta-\alpha\right) + W_0\left(\beta + \alpha\right) - 2\cos\left(4\Im\{\alpha^*\beta\} + \phi \right) W_0(\beta) \Bigr\}.
    \label{echo_general_wigner_fn}
\end{equation}
If one instead of $\hat S_1$ uses $\hat S_1'$ (\eqnref{S1'_def}), the result is
\begin{equation}
    W_{1'}(\beta) = \frac{1}{2}\Bigl\{ W_0\left[-\ii(\beta-\alpha)\right] + W_0\left[-\ii(\beta + \alpha)\right] - 2\cos\left(4\Im\{\alpha^*\beta\} + \phi + 2|\alpha|^2\right) W_0(-\ii\beta) \Bigr\}.
    \label{S1prime_general_wigner_fn}
\end{equation}
Further, one computes
\begin{equation}
    \hat S^\dagger_2\hat \varPi(\beta) \hat S_2 = \frac{1}{2}\Bigl[\hat\varPi(\beta-\alpha) + \hat\varPi(-\beta-\alpha) - \ee^{\ii (\phi + 4\Im\{\alpha^*\beta\})}\hat D(2\beta) - \ee^{-\ii (\phi + 4\Im\{\alpha^*\beta\})}\hat D(-2\beta)\Bigr]
\end{equation}
and therefore
\begin{equation}
    W_2(\beta) = \frac{1}{2}\left[W_0(\beta-\alpha) + W_0(-\beta-\alpha) - \frac{4}{\pi}\Re\left\{\ee^{\ii(4\Im\{\alpha^*\beta\} + \phi)}\chi_0(2\beta)\right\}\right].
    \label{qcmap_general_wigner_fn}
\end{equation}
Here we have identified the characteristic function \cite{barnett_MethodsTheoretical_2002} of the initial state
\begin{equation}
    \chi_0(\beta) \equiv \tr\{\hat D(\beta)\hat\rho_0\} = \int \dd^2 \gamma\, \ee^{2\ii\Im\{\beta\gamma^*\}}W_0(\gamma).
    \label{initial_char_fn}
\end{equation}

Note that these Wigner functions are normalized because we have implicitly assumed an $|\alpha|$ large enough such that there is no overlap between the first two terms in each of Eqs.~(\ref{qcmap_general_wigner_fn}) and (\ref{echo_general_wigner_fn}) (Section~\ref{s:pulse_sequence_analysis}). For these values of $|\alpha|$, the third term in each equation oscillates rapidly enough to integrate to zero.

We now specialize to the thermal state, i.e.~we set $\hat\rho_0 = \hat\rho_T$. The Wigner function of the thermal initial state is
\begin{equation}
    W_T(\beta) = \frac{2\mathcal{P}}{\pi} \ee^{-2\mathcal{P}|\beta|^2}
    \label{thermal_wf}
\end{equation}
where $\mathcal{P} = (2n_\mathrm{th} + 1)^{-1}$ is the purity of the thermal state. The corresponding characteristic function is
\begin{equation}
    \chi_T(\beta) = \ee^{-|\beta|^2/2\mathcal{P}}.
\end{equation}
Replacing $W_0(\beta)$ and $\chi_T(\beta)$ in \eqnref{qcmap_general_wigner_fn} and \eqnref{echo_general_wigner_fn} gives
\begin{equation}
    W_1(\beta) = \frac{\mathcal{P}}{\pi}\left[\ee^{-2\mathcal{P}|\beta-\alpha|^2} + \ee^{-2\mathcal{P}|\beta+\alpha|^2} - 2\cos\left(4\Im\{\alpha^*\beta\} + \phi\right)\ee^{-2\mathcal{P}|\beta|^2}\right].
    \label{echo_thermal_cat_wigner}
\end{equation}
\begin{equation}
    W_2(\beta) = \frac{1}{\pi}\left[\mathcal{P}\left(\ee^{-2\mathcal{P}|\beta-\alpha|^2} + \ee^{-2\mathcal{P}|\beta+\alpha|^2}\right) - 2\cos\left(4\Im\{\alpha^*\beta\} +\phi\right) \ee^{-2|\beta|^2/\mathcal{P}}\right].
    \label{qcmap_thermal_cat_wigner}
\end{equation}
These are the Wigner functions given in the main text up to a redefinition of the parameter $\phi$ ($\phi \to \phi + \pi$). \eqnref{echo_thermal_cat_wigner} is also obtained from $W_1'(\beta)$ if one inserts the thermal Wigner function and redefines $\phi$ to absorb the geometric phase.

\subsection{\label{s:coherence_function}Hot cat state coherence functions}
In this section, we give the full two-dimensional hot cat coherence functions as well as their derivations. As stated in the main text, the first-order coherence function of a quantum state with density matrix $\hat\rho$ is defined as
\begin{equation}
    g(x_1,x_2) \equiv \frac{|\bra{x_1}\hat\rho \ket{x_2}|}{\sqrt{\bra{x_1}\hat\rho\ket{x_1}\bra{x_2}\hat\rho\ket{x_2}}}.
    \label{coherence_function}
\end{equation}
Here $x_{1,2}$ are real-valued dimensionless numbers, and $\ket{x_{1,2}}$ are eigenkets of the quadrature operator
\begin{equation}
    \hat x \equiv \frac{\hat c + \hat c^\dagger}{\sqrt 2},
\end{equation}
meaning that $\hat x \ket {x_1} = x_1 \ket{x_1}$ and equivalently for $x_2$. Due to the positive semidefiniteness of $\hat\rho$, the coherence function is bounded: $1 \geq g(x_1,x_2) \geq 0$. 

We compute the coherence function from $W_{1,2}(\beta)$ using the relation
\begin{equation}
    \bra{x_1} \hat\rho \ket{x_2} = \frac{1}{2}\int_{-\infty}^\infty\dd p\, W\left(\frac{x_1 + x_2}{2\sqrt 2} + \frac{\ii p}{\sqrt 2}\right)\ee^{\ii p (x_1 - x_2)}
    \label{position_element_from_winger_fn}
\end{equation}
which can be derived e.g.~from the expression for computing expectation values from the Wigner function $\langle \hat A\rangle = \tr\{\hat A \hat\rho\} = \int \dd^2\gamma\, 2\tr\{\hat\varPi(\gamma)\hat A\}W(\gamma)$ with $\hat A = \ket{x_2}\bra{x_1}$. It is helpful to first remind ourselves of the coherence function of a thermal state. Using \eqnref{thermal_wf} in \eqnref{position_element_from_winger_fn}, we find the coherence function
\begin{equation}
    g_T(x_1,x_2) \equiv \exp\left\{-\frac{(x_1-x_2)^2}{4}\frac{(1 - \mathcal{P}^2)}{\mathcal{P}}\right\}
    \label{thermal_coherence_function}
\end{equation}
which we denote $g_T$ since it is the coherence function of the thermal state $\hat\rho_T$. The thermal state coherence function is independent of the position on the diagonal $x_1 + x_2$ and is a Gaussian in the distance from the diagonal $x_1 - x_2$. Its standard deviation
\begin{equation}
    \xi_\mathrm{th} \equiv \sqrt{\frac{2\mathcal{P}}{1-\mathcal{P}^2}}.
\end{equation}
is termed the coherence length.
When $\mathcal{P}^2 \ll 1$, $\xi_\mathrm{th} \approx \sqrt{2\mathcal{P}}$. In this case, the coherence length is related to the quadrature standard deviation of the thermal state $\sigma_x \equiv \sqrt{\tr\{\hat x^2 \hat\rho_0\}} = 1/\sqrt{2\mathcal{P}}$ by the reciprocal relationship $\xi_\mathrm{th} = 1/\sigma_x$. In the opposite limit, $\mathcal{P} \to 1$, $\xi_\mathrm{th} \to \infty$ since in this case $g(x_1,x_2) \to 1$. 

We denote the coherence functions of the ECD and qcMAP states respectively as $g_{1}(x_1,x_2)$ and $g_2(x_1,x_2)$. We also introduce the notation $\bar x \equiv (x_1 + x_2)/2$ and $\Delta x \equiv x_1 - x_2$ to make expressions more concise. From \eqnref{coherence_function}, \eqnref{position_element_from_winger_fn} and Eqs. (\ref{qcmap_thermal_cat_wigner}) and (\ref{echo_thermal_cat_wigner}), we compute
\begin{equation}
    g_1(x_1,x_2) = \frac{\ee^{-(\Delta x)^2/2\xi_\mathrm{th}^2}\left|\cosh{(\sqrt 2 \alpha \mathcal{P} 2\bar x)} - \ee^{-4\alpha^2/\xi_\mathrm{th}^2}\cosh{\left(\sqrt 2\alpha\Delta x /\mathcal{P} - \ii\phi\right)}\right|}{\sqrt{\left[\cosh{(2\sqrt 2 \alpha \mathcal{P} x_1)} - \ee^{-4\alpha^2/\xi_\mathrm{th}^2}\cos(\phi)\right]\left[\cosh{(2\sqrt 2 \alpha \mathcal{P} x_2)} - \ee^{-4\alpha^2/\xi_\mathrm{th}^2}\cos(\phi)\right]}}.
    \label{garfield_coherence_exact}
\end{equation}
and
\begin{equation}
    g_2(x_1,x_2) = \frac{\left|\ee^{-(\Delta x)^2/2\xi_\mathrm{th}^2}\cosh\left(\sqrt 2 \alpha \mathcal{P} 2\bar x\right) - \ee^{-(2\bar x)^2/2\xi_\mathrm{th}^2}\cosh\left(\sqrt 2 \alpha\mathcal{P} \Delta x - \ii\phi\right)\right|}{\sqrt{\left[\cosh(2\sqrt 2 \alpha \mathcal{P} x_1) - \ee^{-2x_1^2/\xi_\mathrm{th}^2}\cos(\phi)\right]\left[\cosh(2\sqrt 2 \alpha \mathcal{P} x_2) - \ee^{-2x_2^2/\xi_\mathrm{th}^2}\cos(\phi)\right]}}
    \label{cheshire_coherence_exact}
\end{equation}
These expressions are exact, and we plot them as 2-dimensional functions of $x_1$ and $x_2$ in Figure~S8. In the remainder of this section, we will explain the appearance of Figure~S8 by making approximations.

We are mainly interested in understanding $g_{1,2}$ for $\mathcal{P} \ll 1$ and for $(x_1,x_2)$ away from the origin $(0,0)$ (since our nonzero entries in $\bra{x_1}\hat\rho\ket{x_2}$ are centered around the four points $(\pm \sqrt 2\alpha, \pm \sqrt 2\alpha)$, $(\pm \sqrt 2\alpha, \mp \sqrt 2\alpha)$, and we must have $\alpha > \sigma_x = 1/\sqrt{2\mathcal{P}}$). When the purity is not close to 1, the terms $\propto \cos(\phi)$ in the denominators of $g_{1,2}$ are generally negligible ($g_1$) or nonzero only close to the coordinate axes $(g_2)$. We therefore drop these terms, which makes the denominators of $g_{1,2}$ equal. Using the hyperbolic trig relations, we rewrite the denominators to be $[\cosh{(2 \sqrt 2 \alpha \mathcal{P} 2 \bar x)} + \cosh{(2\sqrt 2 \alpha \mathcal{P} \Delta x )}]/2$. Finally, we approximate $\cosh{(x)} \approx \exp{(|x|)/2}$ for both $\bar x$ and $\Delta x$. This gives
\begin{equation}
    g_1(x_1,x_2) \approx \frac{\exp\left\{-(\Delta x)^2/2\xi_\mathrm{th}^2\right\}}{\sqrt{1 + \exp\left\{2\sqrt 2 \alpha \mathcal{P}(|\Delta x|-2|\bar x|)\right\}}} + \frac{\exp\left\{-(|\Delta x| - 2\sqrt 2\alpha)^2/2\xi_\mathrm{th}^2\right\}}{\sqrt{1 + \exp\left\{2\sqrt 2 \alpha \mathcal{P}(2|\bar x| - |\Delta x|)\right\}}}.
    \label{garfield_coherence_approx}
\end{equation}
\begin{equation}
    g_2(x_1,x_2) \approx \frac{\exp\left\{-(\Delta x)^2/2\xi_\mathrm{th}^2\right\}}{\sqrt{1 + \exp\left\{2\sqrt 2 \alpha \mathcal{P}(|\Delta x|-2|\bar x|)\right\}}} + \frac{\exp\left\{-(2\bar x)^2/2\xi_\mathrm{th}^2\right\}}{\sqrt{1 + \exp\left\{2\sqrt 2 \alpha\mathcal{P}(2|\bar x|-|\Delta x|)\right\}}},
    \label{cheshire_coherence_approx}
\end{equation}
These expressions differ only in the nominator of the last term. The denominators are approximately indicator functions for the $x_1x_2 > 0$ quadrants (first term denominator) and $x_1x_2 < 0$ quadrants (second term denominator) of the $(x_1,x_2)$ plane, i.e.
\begin{equation}
    \frac{1}{\sqrt{1 + \exp\left\{2\sqrt 2 \alpha \mathcal{P}(|\Delta x|-2|\bar x|)\right\}}} \approx [x_1x_2 > 0] = \left\{\begin{array}{cc}
         1 & x_1x_2 > 0 \\
         0 & x_1x_2 < 0
    \end{array}\right.,
    \label{iverson_bracket_approx_1}
\end{equation}
\begin{equation}
    \frac{1}{\sqrt{1 + \exp\left\{2\sqrt 2 \alpha\mathcal{P}(2|\bar x|-|\Delta x|)\right\}}} \approx [x_1x_2 < 0] = \left\{\begin{array}{cc}
         0 & x_1x_2 > 0 \\
         1 & x_1x_2 < 0
    \end{array}\right..
\end{equation}
(The notation $[\ \ ]$ for the conditional expressions is called the Iverson bracket). Using this observation, we arrive at our final expressions for $g_{1,2}$:
\begin{equation}
    g_1(x_1,x_2) \approx g_T(x_1,x_2)[x_1x_2 > 0] + g_T(|x_1-x_2|,2\sqrt 2\alpha)[x_1x_2 < 0].
    \label{g1_indicator_approx}
\end{equation}
\begin{equation}
    g_2(x_1,x_2) \approx g_T(|x_1|,|x_2|).
    \label{g2_indicator_approx}
\end{equation}
When $\mathcal{P} \ll 1$, the approximation \eqnref{iverson_bracket_approx_1} loses accuracy in \eqnref{garfield_coherence_approx}, and in this limit we should instead replace $[x_1x_2 > 0]$ by $1$ in \eqnref{g1_indicator_approx} to get an accurate approximation for $g_1(x_1,x_2)$. Eqs.~(\ref{g1_indicator_approx}) and (\ref{g2_indicator_approx}) agree well with the exact functions plotted in Figure~S8.

Along the line $l(s) \equiv (x_1(s),x_2(s)) = \sqrt 2 \alpha(2s-1,-1)$, $s \in [0,\infty]$ in the $(x_1,x_2)$ plane, $g_1$ and $g_2$ are equal within our approximations leading up to Eqs.~(\ref{g1_indicator_approx}) and (\ref{g2_indicator_approx}). This is the line along which we plot the hot cat state coherence functions in Figure 1C in the main text. 

\newpage
\vspace{1cm}
\begin{center}
    \includegraphics{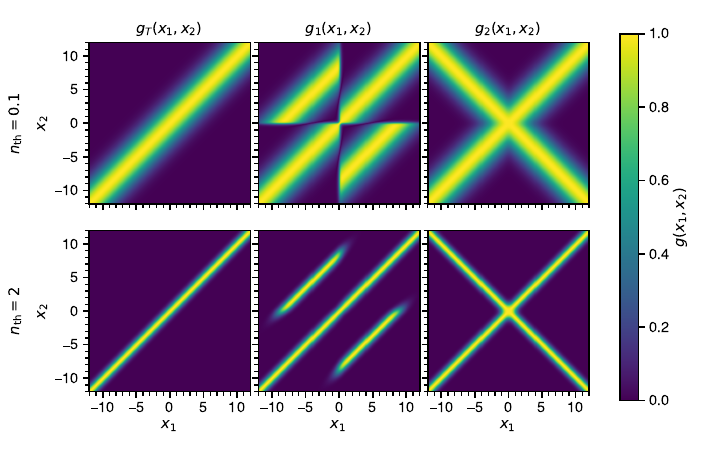}
\end{center}
\noindent {\bf Fig. S8. Hot Schrödinger Cat State Coherence Functions.} Shown is the thermal state coherence function $g_T(x_1,x_2)$ \eqnref{thermal_coherence_function} (left column), and the two cat coherence functions $g_1(x_1,x_2)$ \eqnref{garfield_coherence_exact} (center column) and $g_2(x_1,x_2)$ \eqnref{cheshire_coherence_exact} (right column) for $n_\mathrm{th}$ $0.1$ (top row) and 2 (bottom row). The plot uses $\alpha=3$ and $\phi = 0$.

\vspace{1cm}
\clearpage

\section{Numerical model \label{s:Simulation}}

Based on our characterization of the experimental setup, we introduce the numerical \emph{ab initio} model of our experiment and compare its predictions to the experimentally measured data.

\subsection{Method}
We perform all numerical work using QuTiP version 4.7 \cite{qutip}. We simulate the ECD and qcMAP protocols as follows: Displacement operators and thermal initial states are implemented using QuTiP's built-in functions. The time evolution operations $\hat T(t)$ are implemented using QuTiP's built-in functions to simulate the cavity-qubit dynamics during time evolution, as described in the next paragraph. The qubit and cavity-conditional qubit operations are implemented by simulating the qubit-cavity dynamics under a driving Hamiltonian, as described in more detail below.  This lets us compute the density matrix resulting from the state preparation protocols. We obtain the result of the Wigner function measurement from this density matrix by computing the expectation value of the observable
\begin{equation}
    \hat M(\beta) \equiv \frac{2}{\pi}\hat\varPi(\beta)\left(\ket g \bra g - \ket e \bra e\right).
    \label{measurement_observable}
\end{equation}
This observable gives the outcome of the Wigner function measurement when decoherence and nonlinearities during the measurement sequence are neglected \cite{vlastakis_Sci_342:6158_2013}.

We simulate the time-evolution operators $\hat T(t)$ by using QuTiP's built-in \texttt{mesolve} function to solve the following Lindblad equation
\begin{equation}
    \frac{\partial}{\partial t}\hat\rho(t) = \mathcal{L}\hat\rho(t) \equiv -\frac{\ii}{\hbar}\left[\hat H,\,\hat\rho(t)\right] + \left(\gamma_1\mathcal{D}\left[\ket g \bra e\right] + \frac{\gamma_2}{2}\mathcal{D}\left[\hat \sigma_z\right] + \Gamma\mathcal{D}\left[\hat c\right]\right)\hat\rho(t)
    \label{master_equation}
\end{equation}
from $t=t_0$, where $\hat\rho(t_0)$ is the total cavity-qubit state before the $\hat T$ operation is to be applied, until the final time $t$. The state $\hat\rho(t)$ is then used as input for the next step of the protocol. Here $\gamma_{1} = 1/T_1$, $\gamma_2 = 1/T_2^*$ are the dissipation and dephasing rates of the qubit and $\Gamma = 1/T_{1,\mathrm{c}}$ is the dissipation rate of the cavity. $\mathcal{D}$ is the dissipator superoperator, defined as
\begin{equation}
    \mathcal{D}[\hat A]\hat\rho(t) \equiv \hat A\hat\rho(t)\hat A^\dagger - \frac{1}{2}\left(\hat A^\dagger\hat A \hat\rho(t) + \hat\rho(t)\hat A^\dagger\hat A\right),
\end{equation}
for an arbitrary operator $\hat A$, and $\hat H$ is the Hamiltonian in the interaction picture of the cavity and qubit including the dominant higher-order perturbations
\begin{equation}
    \frac{1}{\hbar}\hat H \equiv - \chi_\mathrm{qc}\hat c^\dagger\hat c\ket e \bra e - \left(\frac{K_\mathrm{c}}{2} + \frac{\chi_\mathrm{qc}'}{2}\ket e \bra e\right)\hat c^\dagger\hat c^\dagger\hat c \hat c.
\end{equation}
We choose the parameters $K_\mathrm{c}, \chi_\mathrm{qc}^\prime, \gamma_1, \gamma_2, \Gamma$ to be the values measured in the experiment (Table S1).

For the qubit operations, we use QuTiP's built-in \texttt{mesolve} function to solve the Lindblad equation
\begin{equation}
    \frac{\partial}{\partial t}\hat\rho(t) = \mathcal{L}\hat\rho(t) - \frac{\ii}{\hbar}\left[\hat H_\mathrm{drive},\,\hat\rho(t)\right]
    \label{driving_lindbladian}
\end{equation}
from time $t_0$, where $\hat\rho(t_0)$ is the total cavity-qubit state before the qubit operation is applied, to time $t$, when the next operation is applied, and we use $\hat\rho(t)$ as initial state for the following operation. The driving Hamiltonian is
\begin{equation}
    \frac{1}{\hbar}\hat H_\mathrm{drive} \equiv \frac{\Omega(t)}{2} \left(\ee^{\ii\phi}\ket e \bra g + \ee^{-\ii\phi}\ket g \bra e\right)
\end{equation}
with
\begin{equation}
    \Omega(t) \equiv \frac{\theta \ee^{-(t-t_0-T/2)^2/2\sigma_t^2}}{\sqrt{2\pi\sigma_t^2}\operatorname{erf}\left(T/2^{3/2}\sigma_t\right)}.
\end{equation}
Here $T$ is the pulse duration so that $t = T + t_0$, $\operatorname{erf}$ is the error function, and $\theta$ is the pulse area over the interval $T$:
\begin{equation}
    \int_{t_0}^{t_0+T} \dd\tau\, \Omega(\tau) = \theta.
\end{equation}
We use $T = 4\sigma_t$ as this is the value used in the experiment. We take $\theta$, $\sigma_t$ and $\phi$ as the parameters of the pulse and choose them according to the desired operation to be modelled. In particular, $\sigma_t = 6$ ns for the global qubit operations, and $\sigma_t = 20$ ns for the cavity-selective disentanglement pulse. We compensate the free evolution times for the finite length of the qubit pulses, so that the maximum of the disentanglement pulse occurs at $t = \pi/\chi_\mathrm{qc}$ in both protocols, and in the ECD protocol, the maximum of the qubit echo pulse occurs at $t = \pi/2\chi_\mathrm{qc}$.

\subsection{Comparison of Simulation and Data}
We run our simulations with all parameters of the Lindbladian taking the values that we measure experimentally (Table S1), and the pulse parameters being the experimentally used values. We present a comparison of the simulation results to the experimental Wigner map data in Figure S9. Here, we simulated with $n_\mathrm{th} = 3.48$, $\alpha = 3.06$ for the ECD protocol and $\alpha = 3.47$ for the qcMAP protocol, and $\phi = \pi$. To facilitate a comparison of the simulations to the data, we displace the simulated Wigner functions so that their fringe pattern aligns with that of the data. We do this by identifying the coherence fringe in the simulated Wigner functions corresponding most closely to the coherence fringe centered at $\beta = 0$ in the data. We then displace the simulated Wigner functions so that the identified fringe is also centered at $\beta = 0$. Specifically, the simulated ECD Wigner function was displaced by $0.654\ii$ and the simulated qcMAP Wigner function was displaced by $0.111 - 0.353\ii$. We additionally rotate the qcMAP state clockwise by $0.163$ radians so that the centers of the displaced thermal states lie along the $\Re\{\beta\}$ axis. The ECD state was not rotated.

Using the same simulation parameters as for the Wigner maps (including the final displacement and rotations), we also simulate the Wigner function linecuts along the $\Re\{\beta\}$ and $\Im\{\beta\}$ axes for the values of $n_\mathrm{th}$ that were reported in Figure 2D-G. We present the computed linecuts in Figure~S10.

\newpage
\begin{center}
    \noindent\makebox[\textwidth]{%
        \includegraphics{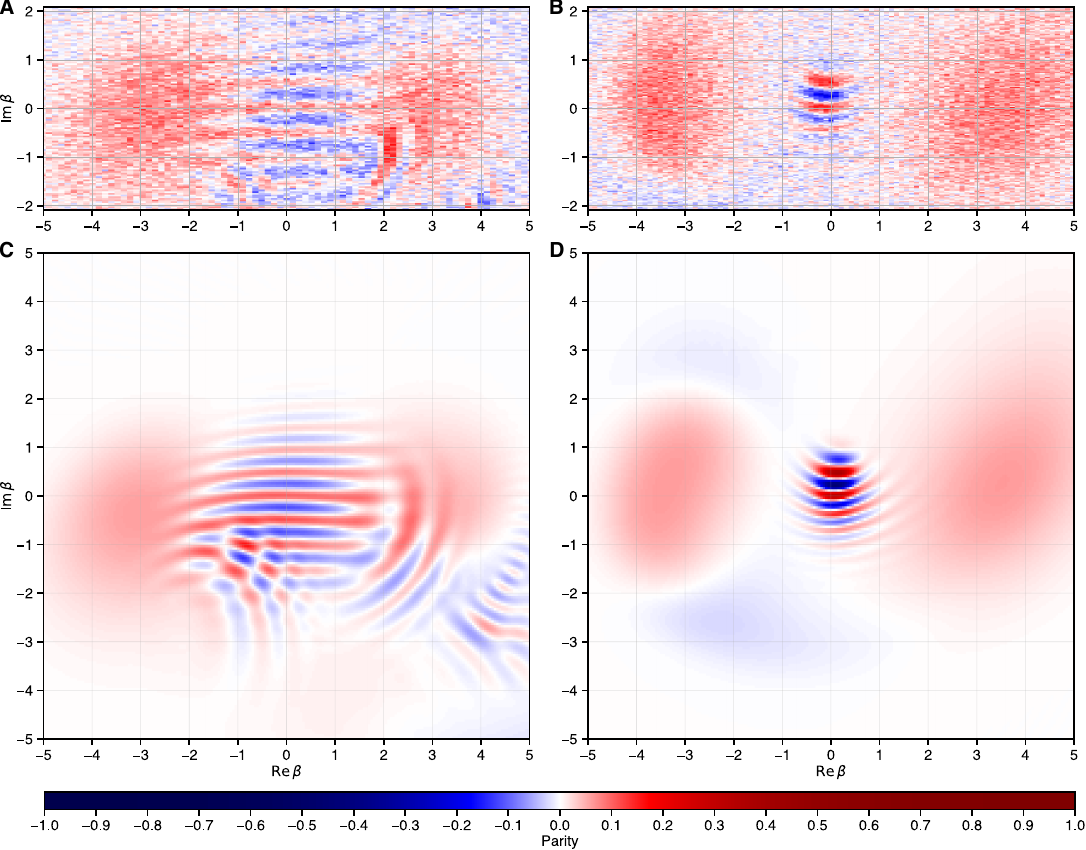}
    }
\end{center}
\noindent {\bf Fig. S9. Comparison of simulated Wigner maps to measured data.} (\textbf{A}) Experimental data obtained for the ECD protocol (also displayed in Figure 1B). (\textbf{B}) Experimental data obtained for the qcMAP protocol (also displayed in Figure 1C). (\textbf{C}) Result of the numerical simulation of the ECD protocol. (\textbf{D}) Result of the numerical simulation of the qcMAP protocol.
\newpage
\clearpage
\begin{center}
    \noindent\makebox[\textwidth]{%
        \includegraphics{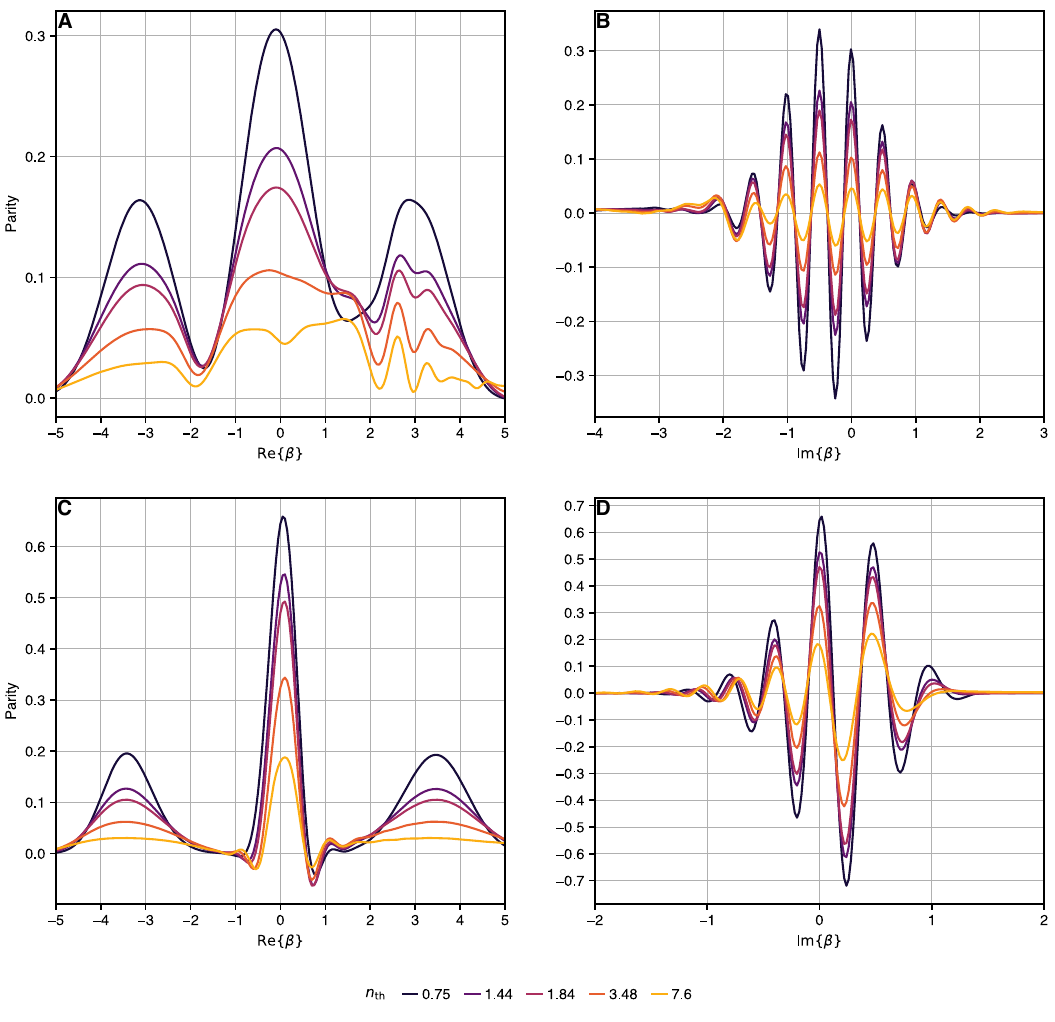}
    }
\end{center}
\noindent {\bf Fig. S10. Wigner function linecuts obtained from simulation.} The line colors are chosen in analogy to the line colors in Figure 1D-G. (\textbf{A}) Linecut along $\Re\{\beta\}$ through the ECD Wigner function. (\textbf{B}) Linecut along $\Im\{\beta\}$ through the ECD Wigner function. (\textbf{C}) Linecut along $\Re\{\beta\}$ through the qcMAP Wigner function. (\textbf{D}) Linecut along $\Im\{\beta\}$ through the qcMAP Wigner function.
\newpage

\subsection{Cavity phase noise}
As discussed in the main text and in Section~\ref{s:parity_lifetime}, the experiment has additional loss channels not included in the numerical model, such as cavity phase noise. In Figure 4 of the main text, we have illustrated the effect of adding this noise to the model by adding a term $\frac{\Gamma_{\phi, c}}{2}\mathcal{D}[\hat n]\hat\rho$ to \eqnref{master_equation}. The dotted curves in Figure 4 are the results of running the numerical model with $\Gamma_{\phi, c} = 1/(80\ \text{µs})$. We stress that this value has been chosen to illustrate the effect, the exact dephasing rate has to be determined in additional experiments. Cavity phase noise is included only in the dotted curves in Figure 4 and nowhere else in this work.

\section{Analysis of Imperfections \label{s:simulation_imperfections}}
The experimentally achieved protocols differ from the ideal ECD and qcMAP protocols analyzed theoretically in Section~\ref{s:theory}. In this section, we study the identified differences in isolation using a combination of analytical and numerical methods. By comparing the results to the ideal protocols and the data, we can attribute features in the data which are not seen in the theoretical Wigner functions to particular experimental imperfections.

\subsection{\label{s:measurement_theory_model}Residual cavity-qubit entanglement}
The experimental imperfections lead to a finite probability $p_e$ of the qubit being in the excited state at the end of the protocol. From \eqnref{measurement_observable} and \eqnref{total_cavit_qubit_state}, one computes that the expected result of the measurement is
\begin{equation}
    W_\mathrm{meas.}(\beta) \equiv \tr\left\{\hat M(\beta)\hat U\hat\rho_0\ket g\bra g\hat U^\dagger\right\} = p_g W_g(\beta) - p_e W_e(\beta).
    \label{Wmeas_def}
\end{equation}
 Here, we have introduced the qubit-conditional Wigner functions
\begin{equation}
    W_{g,e}(\beta) \equiv \frac{2}{\pi}\tr\{\hat \varPi(\beta)\hat\rho_{g,e}\}.
\end{equation}
Imperfect disentanglement between the cavity and qubit will lead to $p_e > 0$, $p_g < 1$. In this case, the output of the Wigner function measurement is not the final state Wigner function, but rather the weighted difference of the Wigner functions of the qubit-conditional states $\hat\rho_g$ and $\hat\rho_e$, with the weights being $p_g$ and $p_e$. As long as $p_g \gg p_e$ and/or the Wigner functions $W_g(\beta)$ and $W_e(\beta)$ do not overlap significantly, the residual qubit-cavity entanglement is a perturbative imperfection to measurement of the final cavity state Wigner function.

\subsection{\label{s:sim_comparison}Comparison of simulations with different parameters}

To understand the effect of each difference between our experiment and the ideal ECD and qcMAP protocols, we run our simulations with different sets of parameters which are chosen to isolate each difference. We present the results of these simulations in Figures S11 (ECD) and S12 (qcMAP). We go through the parameters used for each panel of these figures in the next paragraphs.

`Experimental parameters' (panel A) refers to the simulations as described in Section~\ref{s:Simulation}, where we chose the parameters to match the experiment as closely as possible (however, in Figures S11 and S12, we do not rotate or displace the Wigner functions before plotting).

`Reference parameters' (panel B) refers to a set of parameters chosen to match the ideal ECD and qcMAP protocols considered in Section~\ref{s:pulse_sequence_analysis}. These simulations have no Kerr nonlinearities, negligible width of the non-selective qubit pulses, infinite coherence times, and a disentanglement pulse width which was optimized by sweeping $\sigma_t$ for these parameters and choosing the $\sigma_t$ that minimized $p_e$ in this scenario. Specifically, the reference parameters are: $K_\mathrm{c} = \chi'_\mathrm{qc} = 0$, $\Gamma = \gamma_1 = \gamma_2 = 0$, disentanglement pulse $\sigma_t = 6$ ns, other qubit pulses $\sigma_t = 10^{-13}$ s.

Panels C-F show simulations using the reference parameters but with a subset of the parameters set to their experimental values, as explained in the figure captions.

\clearpage
\begin{center}
    \noindent\makebox[\textwidth]{%
    \includegraphics{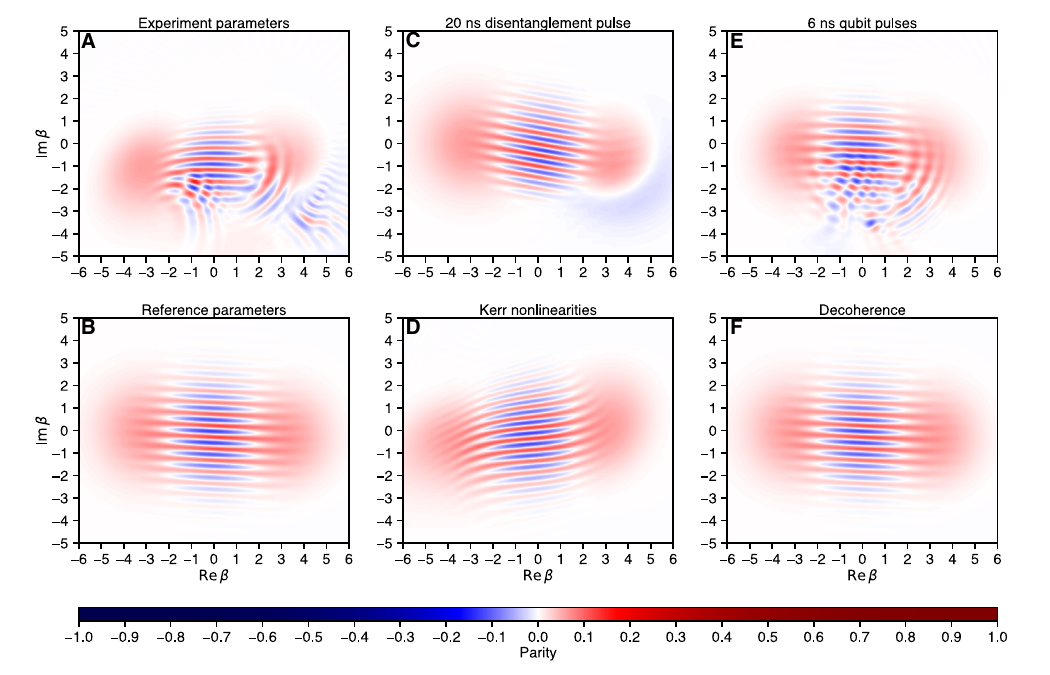}
    }
\end{center}
\noindent {\bf Fig. S11. Comparison of simulations with different parameters for ECD.} We simulate the hot cat state preparation protocols with different sets of parameters, which are chosen to isolate the differences between our experiment and the ideal ECD protocol. All simulations shown used $n_\mathrm{th}=3.48$ and $\alpha=3.06$. (\textbf{A}) Simulation with all parameters taking the values measured or used in the experiment, as in Fig.~S9C. (\textbf{B}) Simulation with parameters chosen to match the ideal ECD protocol (see Section~\ref{s:sim_comparison}). (\textbf{C}) Simulation with the reference parameters but the disentanglement pulse standard deviation $\sigma_t$ set to the experiment value of 20 ns. (\textbf{D}) Simulation with the reference parameters but the Kerr nonlinearity parameters $K_\mathrm{c}$ and $\chi_{\mathrm{qc}}'$ taking their experimentally measured values. (\textbf{E}) Simulation with the reference parameters but the $\sigma_t$ of all qubit pulses set to 6 ns, the minimum pulse $\sigma_t$ that our experimental instrumentation can achieve. (\textbf{F}) Simulation with the reference parameters but with the coherence times $T_1$, $T_2^*$ and $T_{1,\mathrm{c}}$ taking their experimental values.
\vspace{1cm}
\newpage
\begin{center}
    \noindent\makebox[\textwidth]{%
    \includegraphics{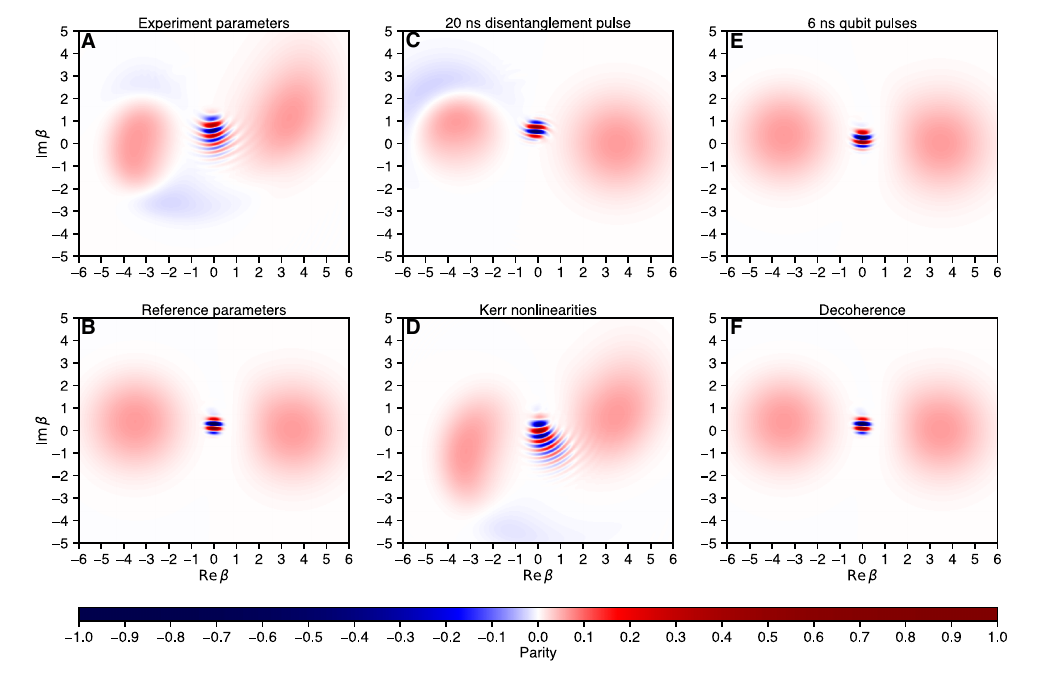}
    }
\end{center}
\noindent {\bf Fig. S12. Comparison of simulations with different parameters for qcMAP.} We simulate the hot cat state preparation protocols with different sets of parameters, which are chosen to isolate the differences between our experiment and the ideal qcMAP protocol. All simulations shown used $n_\mathrm{th}=3.48$ and $\alpha=3.47$. (\textbf{A}) Simulation with all parameters taking the values measured or used in the experiment, as in Fig.~S9D. (\textbf{B}) Simulation with parameters chosen to match the ideal qcMAP protocol (see Section~\ref{s:sim_comparison}). (\textbf{C}) Simulation with the reference parameters but the disentanglement pulse standard deviation $\sigma_t$ set to the experiment value of 20 ns. (\textbf{D}) Simulation with the reference parameters but the Kerr nonlinearity parameters $K_\mathrm{c}$ and $\chi_{\mathrm{qc}}'$ taking their experimentally measured values. (\textbf{E}) Simulation with the reference parameters but the $\sigma_t$ of all qubit pulses set to 6 ns, the minimum pulse $\sigma_t$ that our experimental instrumentation can achieve. (\textbf{F}) Simulation with the reference parameters but with the coherence times $T_1$, $T_2^*$ and $T_{1,\mathrm{c}}$ taking their experimental values.
\newpage

\subsection{Free evolution timing errors}
In the experiment, all cavity and qubit operation pulses take a finite time. Cavity displacement pulses take 16 ns, and qubit operations take $4\sigma_t$ to complete (24 ns for non-selective qubit pulses, 80 ns for the disentanglement pulse). The design of the experimental protocol takes this into account by optimizing the pulse timings, with a step size of 4 ns. Nevertheless, here we theoretically investigate the effects of perturbing the time argument of the free evolution operator $\hat T(t)$ in the qcMAP protocol. We show that such perturbations, if present, lead to bending distortions of the hot cat state fringes which are not seen for cold cat states. 

In the ideal qcMAP protocol, the free evolution time is $\pi/\chi_\mathrm{qc}$. Here we take the free evolution time to be $\pi/\chi_\mathrm{qc} + \tau$, where $\tau$ is the timing error. Following the analysis in Section~\ref{s:pulse_sequence_analysis}, the effective operator of the protocol becomes
\begin{equation}
    \hat S_1 = \frac{1}{\sqrt{2}} \left[1 - \exp\left\{\mathrm{i}\left(\chi_\mathrm{qc} \tau \hat n + \phi\right)\right\} \hat \varPi \right]\hat D(\alpha).
\end{equation}
The Wigner function resulting from the application of this operator on a cavity state can be computed using the same methods as in Section~\ref{s:general_wigner_functions}. The result is
\begin{multline}
    W_1(\beta) = \frac{1}{2}\biggl[W_0(\beta-\alpha) + W_0(-\alpha-\beta\ee^{-\ii\chi_\mathrm{qc}\tau}) \\
    - \frac{4}{\pi}\Re\left\{\ee^{\ii\varphi}\tr\left\{\hat D\left[2\beta + \alpha(\ee^{\ii\chi_\mathrm{qc}\tau} - 1)\right]\exp\left\{\ii\chi_\mathrm{qc}\tau\hat n\right\}\hat\rho_0\right\}\right\}\biggr]
    \label{timing_error_wigner_fn}
\end{multline}
with
\begin{equation}
    \varphi = \phi + 2\Im\left\{\alpha^*\beta\left[1+\exp\left(-\ii\chi_\mathrm{qc}\tau\right)\right]\right\} + |\alpha|^2\sin\left(\chi_\mathrm{qc}\tau\right).
    \label{timing_error_phase}
\end{equation}
For a thermal state $\hat\rho_0 = \hat\rho_T$,
\begin{multline}
    \tr\left\{\hat D\left[2\beta + \alpha(\ee^{\ii\chi_\mathrm{qc}\tau} - 1)\right]\exp\left\{\ii\chi_\mathrm{qc}\tau\hat n\right\}\hat\rho_0\right\} = \\
    = \frac{1}{1 + n_\mathrm{th}\left(1-\ee^{\ii\chi_\mathrm{qc}\tau}\right)}\exp\left\{-\left(\frac{1}{2} + \frac{n_\mathrm{th}\ee^{\ii\chi_\mathrm{qc}\tau}}{1 + n_\mathrm{th}\left(1-\ee^{\ii\chi_\mathrm{qc}\tau}\right)}\right)\left|2\beta + \alpha\left(\ee^{\ii\chi_\mathrm{qc}\tau} - 1\right)\right|^2\right\}.
    \label{timing_error_trace}
\end{multline}
Taking $\tau = 0$ recovers \eqnref{qcmap_thermal_cat_wigner} as expected. To understand the effect of a small $\tau$, we linearize \eqnref{timing_error_phase} and \eqnref{timing_error_trace} in $\tau$. Taking $\alpha$ to be real, the coherence term in \eqnref{timing_error_wigner_fn} linearized in $\tau$ is
\begin{multline}
    \frac{4}{\pi}\cos\left\{4\alpha \Re\{k^* \beta\} + \phi' + 4\chi_\mathrm{qc}\tau |\beta|^2n_\mathrm{th}(1 + n_\mathrm{th})\right\}\\ \cdot \exp\left\{-2(2n_\mathrm{th} + 1)\left(|\beta|^2 + \chi_\mathrm{qc}\tau\alpha\Im\{\beta\}\right)\right\}.
\end{multline}
Here $k \equiv - \chi_\mathrm{qc}\tau / 2 + \ii$ and $\phi' \equiv \phi + \chi_\mathrm{qc}\tau(|\alpha|^2 + n_\mathrm{th})$.
This expression contains three additional effects compared to the $\tau=0$ case: 1) The phase shift has changed from $\phi$ to $\phi'$. 2) Since the left Gaussian has moved in the $\Im\{\beta\}$ direction, the center of the fringes has also moved in the $\Im\{\beta\}$ direction, and $k$ has obtained a real part. 3) The presence of $|\beta|^2$ in the cosine argument causes a bending distortion of the fringes. The $|\beta|^2$ term vanishes when $n_\mathrm{th}\to 0$, so that noticeable bending of the fringes occurs only for hot cats. 

For illustration, we plot \eqnref{timing_error_wigner_fn} in Figure~S13 for $\alpha=3.47$, $n_\mathrm{th}=3.48$, $\phi=\pi$, and a timing error of $\tau = 20$ ns. We emphasize that this value of $\tau$ is much larger than the 4 ns steps with which we optimize the pulse timings in our experiment. Figure S13 bears some resemblance to Figures S9 and S10; however, our numerical simulations do not contain any timing error, and the fringe bending observed in our experiment is mainly due to the Kerr nonlinearity (see Section~\ref{s:sim_comparison}).

\newpage
\begin{center}
    \noindent\makebox[\textwidth]{%
        \includegraphics{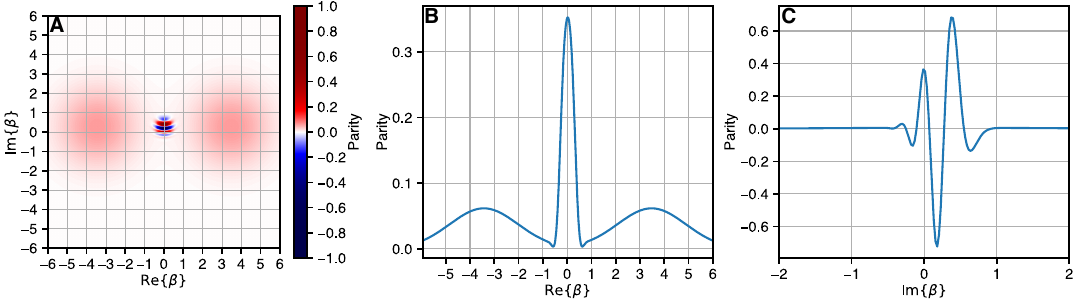}
    }
\end{center}
\noindent {\bf Fig. S13. Fringe bending due to timing errors.} (\textbf{A}) Wigner function \eqnref{timing_error_wigner_fn} for a timing error of 20 ns. The Wigner function has been rotated and displaced before plotting to remove the extra rotation due to the timing error and center the fringe pattern at $\beta = 0$. (\textbf{B}) Linecut through the Wigner function in panel A along $\Re{\beta}$. (\textbf{C}) Linecut through the Wigner function in panel A along $\Im{\beta}$.
\newpage

\end{document}